\documentclass[aps, superscriptaddress,amsmath,amssymb,floatfix, notitlepage, bm,braket, twocolumn]{revtex4-1}
\usepackage{times}
\usepackage{graphics}
\usepackage{graphicx}
\usepackage{color}
\usepackage{bm}
\usepackage{braket}
\usepackage{mathtools}
\usepackage{mathcomp}
\usepackage[caption=false,justification=raggedright,position=top,singlelinecheck=off]{subfig}
\usepackage{soul}
\usepackage{tabularx}
\usepackage{calrsfs}
\usepackage[vcentermath]{youngtab}
\usepackage[mathscr]{euscript}

\newcommand{\dn}{\downarrow}

\newcommand{\up}{\uparrow}

\newcommand{\enni}{\noindent}
\newcommand{\enbe}{\begin{equation}}
\newcommand{\enee}{\end{equation}}
\newcommand{\enba}{\begin{align}}
\newcommand{\enea}{\end{align}}

\begin{document}

\title{Interacting Majorana fermions in strained nodal superconductors}
\author{Emilian M.\ Nica}
\author{Onur Erten}
\affiliation{Department of Physics
Box 871504
Arizona State University
Tempe, Arizona  85287-1504}
\email[Corresponding author: ]{enica@asu.edu}
\date{\today}


\begin{abstract}
Landau levels (LL) have been predicted to emerge in systems with Dirac nodal points under applied non-uniform strain. We consider 2D, $d_{xy}$ singlet (2D-S) and 3D $p \pm i p$ equal-spin triplet (3D-T) superconductors (SCs). We demonstrate the spinful Majorana nature of the bulk gapless zeroth-LLs. Strain along certain directions can induce two topologically distinct phases in the bulk, with zeroth LLs localized at the the interface. These modes are unstable toward ferromagnetism for 2D-S cases. Emergent real-space Majorana fermions in 3D-T allow for more exotic possibilities. 
\end{abstract}

\maketitle

Strain engineering has emerged in recent years as a promising way to access topologically non-trivial phases. The initial proposal of Landau levels (LLs) in strained graphene~\cite{Guinea} was followed by a number of extensions, most notably to nodal Dirac superconductors (SCs)~\cite{Nica, Massarelli} and Weyl semi-metals and SCs~\cite{Grushin, Liu}. These intriguing proposals raise questions regarding the nature and the stability of the strain-induced gapless modes. Such issues have not been addressed in detail except in the case of strained graphene\cite{Ghaemi}. Here, we study the bulk gapless LLs which emerge under applied non-uniform strain in prototypical models which describe a large class of well-studied 2D d-wave SCs, as exemplified by high-$T_{c}$ cuprates~\cite{Hashimoto}. Furthermore, we consider possible instabilities driven by small residual interactions in the low-energy sector. We likewise analyze 3D time-reversal symmetric equal-spin triplet SCs.

We first elucidate the general spinful Majorana nature of the bulk zeroth LLs which satisfy 

\enni \begin{align}
\gamma_{s0}(\bm{k}) = \sum_{s'} M_{ss'} \gamma^{\dag}_{s'0}(-\bm{k}),
\label{Eq:Mjrn}
\end{align}

\enni where $s= \up/\dn$ is a spin index with $M= -\sigma_{y}$ for 2D singlet (2D-S) SCs with $d_{xy}$ pairing. Similarly, $M= -i \sigma_{z}$ for 3D triplet (3D-T) with $p \pm i p$ pairing. Subsequently, we show that the topological properties of the strained bulk depend on the direction of the uniaxial strain. When non-uniform strain is applied along an axis of the Brillouin Zone (BZ), a single zeroth LL per spin $s$ naturally emerges as a consequence of a phase transition in the bulk which separates two topologically distinct gapped regions. Whenever two independent zeroth LLs per spin are present, as in the case of 2D-S SCs with strain along a diagonal of the BZ, the bulk is topologically trivial. Furthermore, we show that the instabilities of these strained systems depend on the Majorana properties of the zeroth LLs. In 2D-S systems, the halving of the degrees-of-freedom (DOF) associated with the zeroth LLs effectively excludes any valley-polarization instabilities, unlike the case in graphene \cite{Ghaemi}. Instead, the zeroth LLs in flat bands are generically unstable to ferromagnetism induced by small residual short-range interactions. By contrast, we show that the strained 3D-T SCs exhibit emergent \emph{real-space} Majorana gapless modes in the bulk.  Generic density-density interactions are suppressed in these cases, but non-local spin-exchange four-fermion terms of the form 
$U(\bm{R}, \bm{R'}) c_{P,\sigma}(\bm{R}) c_{P,\bar{\sigma}}(\bm{R}) c_{P,\sigma}(\bm{R}')c_{P,\bar{\sigma}}(\bm{R}')$ are allowed and can lead to more exotic instabilities. The real-space Majorana fermions $c_{P, \sigma}(\bm{R})$ arise upon projection onto the zeroth LL sector. Our conclusions are supported by detailed analytical calculations, which are available in the Supplementary Materials (SM), as well as by numerical results presented further down.

\begin{figure}[t!]
\includegraphics[width=0.75\columnwidth]{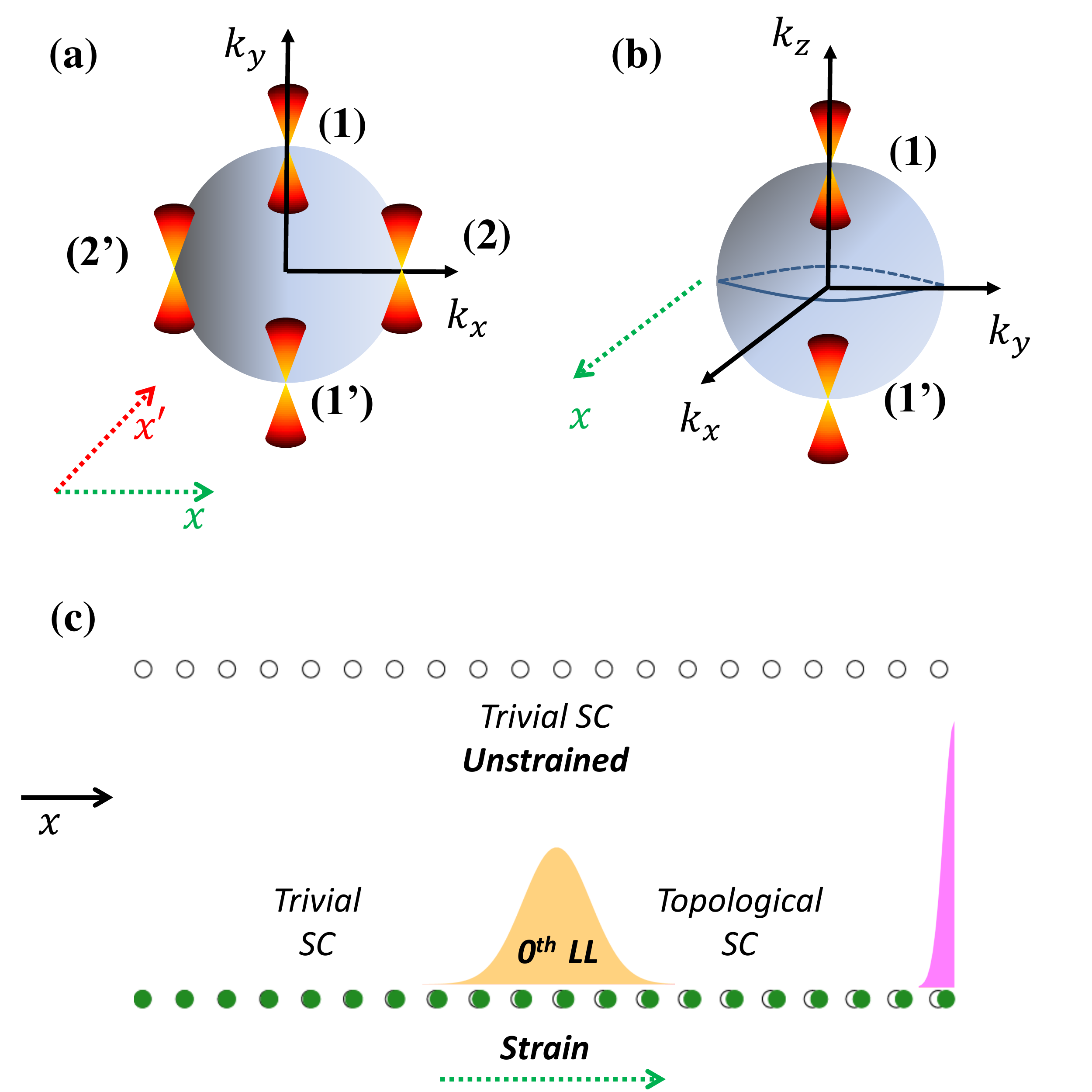}
\caption{Nodal Dirac spectrum and Fermi surfaces for (a) 2D $d_{xy}$ pairing and (b) 3D $(p\pm ip)$. Note the labeling of the valleys. (c) The green and red allows indicate two directions of applied uniaxial strain along the axes and diagonal directions, respectively. (c) Emergence of the zeroth LLs at the boundary between strain-induced topologically distinct phases in the bulk.}
\label{Fig:1}
\end{figure}

\noindent
\emph{Spinful Majorana fermions -} The pairing Hamiltonian under applied uniaxial strain along the $x$ direction can be written as 

\enni \begin{align}
H= \sum_{\bm{k}} \bm{\Psi}^{\dag}(\bm{k}) \left[ \hat{h}(\bm{k}) + \hat{\Delta}(\bm{k}) \right] \bm{\Psi}(\bm{k}),
\end{align}

\enni where $\Psi^{T}(\bm{k})= \left(c_{\sigma}(x_{i}, \bm{k}), (i \sigma_{y})_{\sigma \sigma'} c^{\dag}_{\sigma}(x_{i}, -\bm{k}) \right)$. The index $i$ runs over all $N_{x}$ lattice sites along the $x$ direction. The conserved momentum $\bm{k}$ is perpendicular to the direction of the applied strain. The normal part given by the spin-independent $\hat{h} \sim \sigma_{0}\tau_{z}$ contains the effects of strain. The pairing part is given by $\hat{\Delta} =\hat{\Delta}_{S} \sigma_{0} \tau_{x}, \hat{\Delta}_{T, x} \sigma_{x}\tau_{x} +  \hat{\Delta}_{T, y} \sigma_{y} \tau_{x}$, for  2D-S and 3D-T cases, respectively. Here, the $\sigma$ and $\tau$ represent Pauli matrices in spin and Nambu spaces, respectively. The Hamiltonian represents a set of effective 1D chains defined for each conserved momentum $\bm{k}$. For more details, we refer the reader to the SM.

In the absence of any strain, the low-energy solutions of the Hamiltonian determine four nodal points on a generic Fermi surface (FS) for the 2D-S and two nodal points for 3D-T cases. The resulting spectra are illustrated in Figs.~\ref{Fig:1}~(a) and (b), respectively.

By analogy with graphene~\cite{Guinea}, weak and non-uniform strain plays the role of an effective vector potential in the low-energy theory and the spectrum exhibits LLs. As pointed out in Ref.~\onlinecite{Nica}, the strained system is time-reversal symmetric. Consequently, the  pseudo-magnetic fields are not subject to Meissner screening, in contrast to the case of genuine magnetic fields where LLs are suppressed~\cite{Franz}.

In either 2D-S or 
3D-T cases, the Bogoliubov-de Gennes (BdG) Hamiltonian can be decoupled into two independent spin sectors $H_{s}(\bm{k})$. To illustrate, in the 2D-S case the two sectors involve $c_{\up}(x_{i}, \bm{k}), c^{\dag}_{\dn}(x_{i}, -\bm{k})$ and their spin-flipped counterparts, respectively. In the 3D-T case, the two sectors involve fermions of equal spin. The BdG equations for each spin sector are $H_{s}( \bm{k}) \Psi_{ns}( \bm{k}) = E_{ns} \Psi_{ns}( \bm{k}), $
where $\Psi^{T}_{ns} = \left(u_{ns}(x_{i}, \bm{k}), v_{ns}(x_{i}, \bm{k}) \right)$ is a BdG spinor, $s$ is the spin of the BdG quasiparticles, and $n$ is a LL index. We define particle-hole (p-h) transformations~\cite{Sato} for singlet and triplet pairing as $C_{S} = (i\tau_{y}) \otimes (\bm{1})_{N \times N} K$ and $C_{T}= \tau_{x} \otimes (\bm{1})_{N \times N} K$, respectively, where $K$ represents complex-conjugation. Since $C H_{s}(\bm{k})C^{-1} = -H_{s}(-\bm{k})$, it follows that $H_{s}(\bm{k})$ has eigenstates $\Psi_{sn}(\bm{k})$ and $C\Psi_{sn}(-\bm{k})$ at energies $E$ and $-E$, respectively. 

Whenever $H_{s}(\bm{k})$ exhibits  a single zeroth LL, only one of the two states $\Psi_{s0}(\bm{k})$ and $C\Psi_{s0}(-\bm{k})$ is a non-trivial solution. In practice, only \emph{half} of all momenta for each spin sector are associated with independent DOF. We take this as a definition of spinful Majorana modes. Equivalently, the analytical solutions for all cases (Sec.~II C,~III B, and IV B of the SM) indicate that the BdG quasiparticles in the zeroth LLs are pair-wise related via Eq.~\ref{Eq:Mjrn}. In order to reconcile this redundancy with the BdG transformation, we constrain the zeroth LL sector to half of the allowed momenta. This procedure preserves the correct normalization of the $c_{\sigma}(x_{i}, \bm{k})$ operators, as shown in Sec.~II D of the SM.

\noindent
\emph{Topological origin of zeroth LLs - }
Whenever strain induces a single zeroth LL of spin $s$ and momentum $k_y$ in the bulk, it is located at the interface of two fully-gapped and topologically distinct phases. To illustrate, we consider an effective 1D chain which is gapped in the pristine 2D-S case. Non-uniform strain along one of the axes locally closes the gap in the bulk of the 1D chain, and induces a transition between two topologically distinct phases which gives rise to the lowest LL. This mechanism is illustrated in Figs.~\ref{Fig:1}~(c) and by our numerical results presented further down. A similar picture emerges in the 3D T case. Whenever two independent zeroth LLs per spin $s$ are induced in the bulk by applied strain, the gapped regions on either side are topologically trivial. This occurs in the 2D-S case with strain along a diagonal (Fig.1(b)). 

Thus the zeroth LLs are closely related to topologically nontrivial edge states. Indeed, under applied strain, our effective 1D and 2D models belong to the DIII class of the Altland-Zirnbauer classification~\cite{Teo}. The latter predicts Majorana Kramers doublets~\cite{Teo} for point defects in 1D, while line defects in 2D lead to helical Majorana fermions, both associated with a non-trivial bulk $Z_{2}$ invariant. This is precisely what we obtain in the 2D-S case with strain along an axis of the BZ and in the 3D-T case, respectively. Similarly, the presence of two zeroth LLs in the 2D-S case with strain along a diagonal of the BZ signals topologically trivial bulk phases. 

We stress that the topological transition in the bulk via applied strain does not imply destruction of the pairing ground-state. We argue by analogy to vortex states in standard Type-II SCs, where the condensate likewise survives~\cite{de_Gennes}.

\noindent
\emph{Results - }
We first consider the case of a 2D-S cases. Referring to Fig.~\ref{Fig:1}~(a), the unstrained system exhibits Dirac points on the axes of the BZ, located at $\bm{k} = \left(0, \pm K_{F} \right)$ and labeled by valley indices $(1)$ and $(1')$. The pair of valleys  located at $\bm{k} = \left(\pm K_{F}, 0 \right)$ are labeled by $(2)$ and $(2')$, respectively. We introduce uniaxial strain along the $x$ direction (green arrow in Fig.~\ref{Fig:1}~(a)) via a slowly-varying correction to the hopping term $\delta t(x_{i}) = t\epsilon x_{i}$, where $t$ is the pristine hopping coefficient. As discussed in Sec.~V, the strain parameter $\epsilon$ is proportional to the gradient of the applied strain. The detailed form of the lattice Hamiltonian is given in Sec.~II A of the SM. Next we solve this system analytically in the continuum limit about each of the four valleys. Details of the calculation are presented in Sec.~II B and C of the SM. We find that strain induces LLs at valleys $(1)$ and $(1')$. Furthermore, the zeroth LL states obey Eq.~\ref{Eq:Mjrn}. Therefore, the gapless modes of opposite valley and spin indices are not independent. 
\begin{figure}[t!]
\includegraphics[width=0.75\columnwidth]{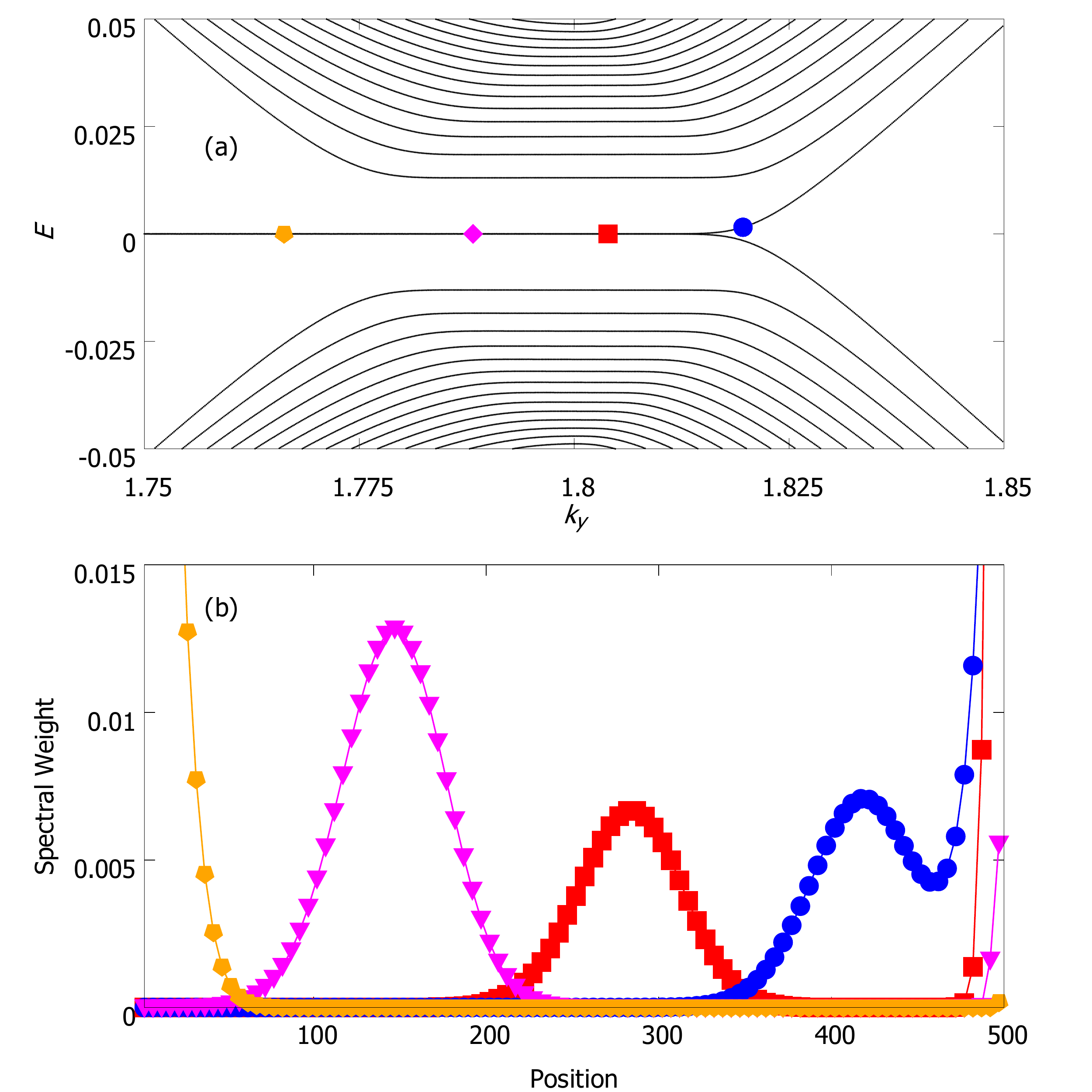}
\caption{(a) Spectrum for the 2D-S case  about valley $(1)$ as a function of conserved momentum and strain along the $x$ direction for the spin up sector. See text for the parameters of the calculation.~(b) Evolution of the spectral weight with momentum for the two degenerate states at zero-energy. Note the formation of a broad LL state in the bulk accompanied by an edge state.}
\label{Fig:2}
\end{figure}
Likewise, the numerical solutions for momenta approaching the Dirac points are consistent with the analytical results. To illustrate, in Fig.~\ref{Fig:2}~(a) we show the spectrum of $H_{s=\up}$ about valley $(1)$. The results were determined using a chain of 500 sites with lattice spacing $a=1$, pairing amplitude $\Delta=0.1$, and strain parameter $\epsilon=9 \times 10^{-4}$. We can distinguish the two zero-energy states via their respective spectral weights. In Fig.~\ref{Fig:2}~(b) we show the spectral weight $|u(x_{i}, k_{y})|^{2} + |v(x_{i}, k_{y})|^{2}$ as a function of position along the effective 1D chain, for the several decreasing values of $k_{y}$ indicated in Fig.~\ref{Fig:2}~(a). As we approach the center of the flat zeroth LL band, the spectral weight shows well-defined peaks in the bulk and at the edge, corresponding to a zeroth LL and an edge state, respectively. This confirms the topological origin of the zeroth LL modes as pointed out in Fig.~\ref{Fig:1}~(c). Indeed, the bulk of the strained sample now contains two topologically-distinct phases, separated from each other by the zeroth LL. Likewise, an edge state emerges at the boundary between the vacuum and the non-trivial sector. As we move away from the node, the zeroth LL merges with an edge state (orange points), separating the topologically trivial vacuum from the non-trivial bulk. These results also illustrate that there is only one bulk, zeroth LL per spin sector, confirming it's Majorana nature. A similar result is obtained for the opposite spin sector and about valley $(1')$.

\begin{figure}[t!]
\includegraphics[width=0.75\columnwidth]{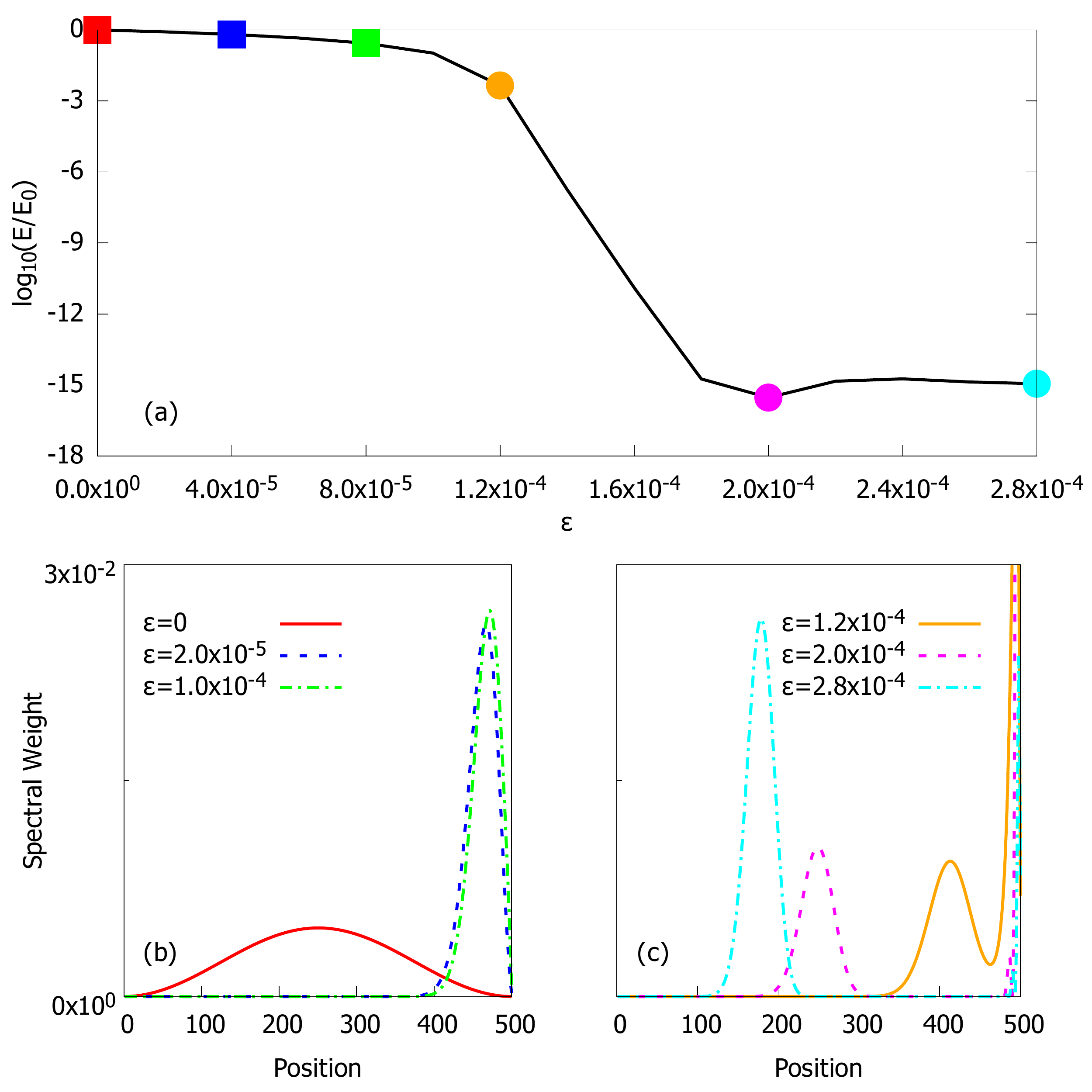}
\caption{2D-S case with strain along axis: (a) suppression of the bulk gap $E_{0}$ for $k_{y}=1.84$ as a function of increasing strain parameter $\epsilon$. Beyond $\epsilon \approx 1.2 \times 10^{-4}$ the gap effectively vanishes.~(b) Evolution of the spatially-resolved spectral weight with increasing strain indicated in (a). Although the vanishing strain cannot close the gap, spectral weight accumulates at the edge.~(c) Beyond a threshold strain $\epsilon=1.2 \times 10^{-4}$, the gap suppression indicated in (a) is accompanied by the emergence of the bulk zeroth LL and edge states. In between the two localized states, the chain is in a topologically non-trivial phase, as illustrated in Fig.~\ref{Fig:1}~(c). The parameters of the calculation are the same as in Fig.~\ref{Fig:2}.}
\label{Fig:3}
\end{figure}

Our hypothesis can be further supported by considering the evolution of the gapped system at $k_{y}=1.84$, as a function of applied strain, as shown in Fig.~\ref{Fig:3}~(a). Here $E_{0} \approx O(10^{-2})$ is the value of the gap in the absence of strain. As the strain parameter increases, the spectral weight is increasingly localized at the edge, as shown in Fig.~\ref{Fig:3}~(b). Upon reaching a strain parameter $\epsilon \approx 1.2\times 10^{-4}$, the gap first decreases dramatically then stabilizes close to zero. The corresponding spectral weights in Fig.~\ref{Fig:3}~(c) exhibit two sharp peaks corresponding to zeroth LL and edge states, respectively. 

Next, we discuss the many-body instabilities of the spinful Majorana fermions. We consider residual short range interactions of the Hubbard type in the paired state \cite{Potter} (see SM Sec. II E). Provided that the cyclotron frequency $\omega_{c} \sim \sqrt{B}$ associated with the pseudo-magnetic field $B$ is much smaller than the interaction strength U', we project the lattice operators onto the zeroth LL sector. Since only half of the zero modes are independent (Eq.\ref{Eq:Mjrn}), we can formally eliminate the valley $(1')$ DOF (See SM Sec. II D). Consequently, we obtain an effective Hubbard interaction involving quasiparticles of both spins at valley $(1)$. At mean-field level, this model predicts an instability toward ferromagnetism with a flat gap of magnitude $n_{0}U'/2$. Here, $n_{0}\sim l^{-2}_{B}$ is the density of zeroth LLs and $l_{B}$ is a pseudo-magnetic length. The remaining possible instabilities are either energetically unfavored, as in the case of singlet pairing, or are excluded due to the halving of the DOF in the case of a spin-density wave. 

Note that LLs do not emerge at valleys $(2)$ and $(2')$. As shown in Sec.~II C of the SM, in the continuum limit, the effects of the strain can be incorporated via a global phase. The resulting states remain gapless and exhibit an approximate linear-in-$q$ dispersion. Using standard RPA arguments, we see that coupling the projected zeroth LLs to the gapless modes at valleys $(2)$ and $(2')$ does not qualitatively change our conclusion. 

\begin{figure}[t!]
\includegraphics[width=\columnwidth]{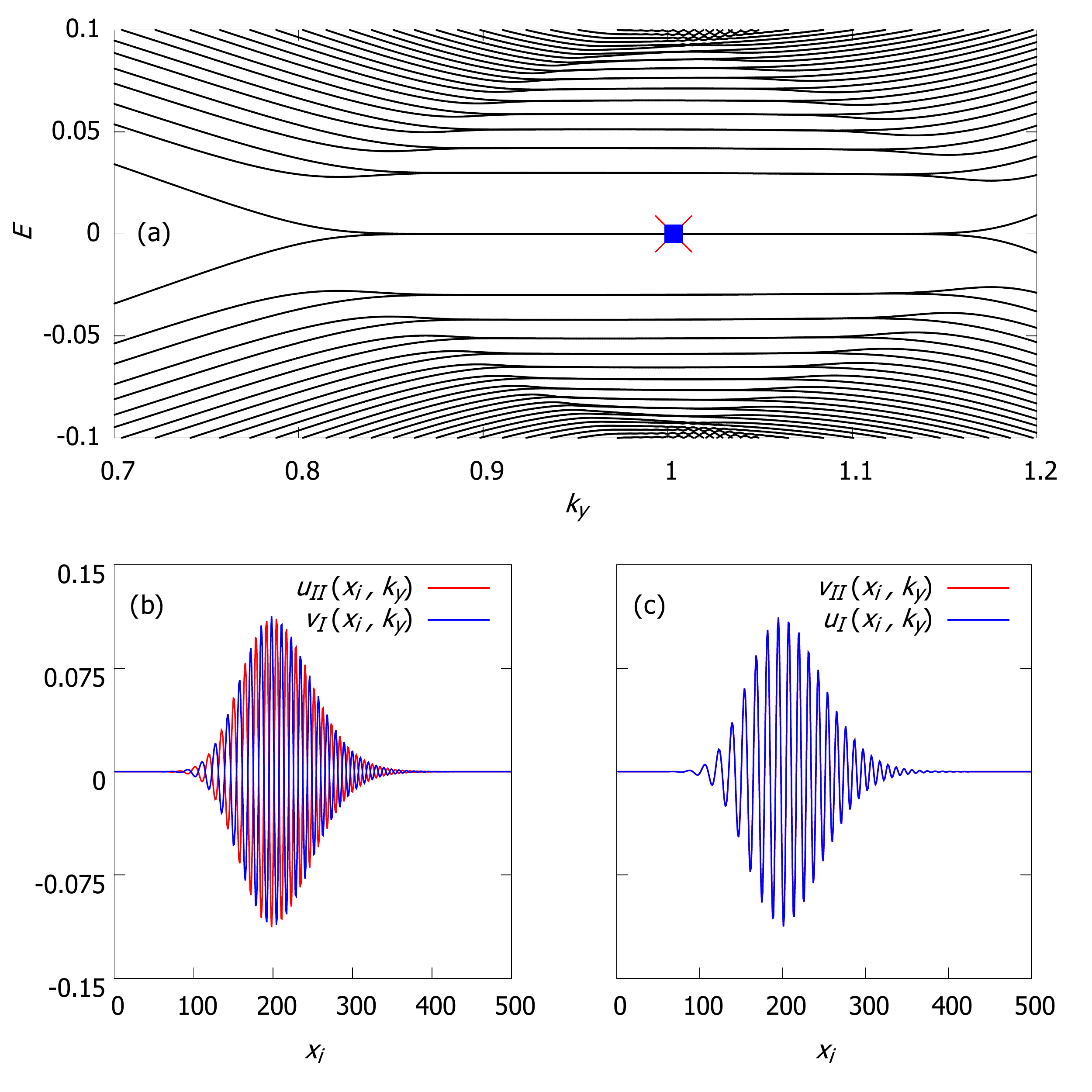}
\caption{2D-S case with strain along the diagonal of the BZ. (a) Spectrum as a function of conserved momentum. The degenerate states at zero energy are both in the bulk. See text for the parameters of the calculation.~(b) Real BdG coefficients as functions of position along the 1D chain for the two states indicated in (a). The two sets of coefficients are related via $u_{II}= -v_{I}, v_{II}=u_{I}$ due to the mirror symmetry as discussed in the main text.}
\label{Fig:4}
\end{figure}

We now consider 2D-S SCs with strain applied along the diagonal of the BZ (red arrow in Fig.~\ref{Fig:1}~(a)). These differ from the analogous cases of strain along an axis of the BZ in several ways. First, there are only two valleys at $k \approx \pm K_{F}$, where the conserved momentum is perpendicular to the diagonal of the BZ. Secondly, the 1D effective models are defined on two sublattices. Moreover, the Hamiltonians are invariant under a reflection $H_{s}(k)=H_{s}(-k)$. Together with the p-h symmetry discussed previously, it ensures that zero-energy modes of the same spin and at the same valley emerge in pairs with BdG coefficients $\Psi^{T}_{s}(k)= \left( u_{s}(x_{i}, k), v_{s}(x_{i}, k) \right)$ and $C \Psi_{s}(k)=(-v_{s}(x_{i}, k), u_{s}(x_{i}, k))$. Note the absence of the minus sign for the momentum of the second spinor. For more details, please consult Sec.~III A of the SM.

A consequence of the combined mirror and p-h symmetry is that there are two independent zeroth LLs per valley and spin, as illustrated by analytical solutions of the BdG equations in the continuum limit (Sec.~III B of the SM). The numerical solutions of the effective 1D lattice model likewise support this claim. In Fig.~\ref{Fig:4}, we present our numerical results for a chain of 500 sites with pairing amplitude $\Delta=10^{-1}$ and strain parameter $\epsilon=5\times 10^{-3}$. The BdG coefficients of the two degenerate zero-modes in panel (a) are precisely of the form imposed by the enhanced symmetry, as indicated in panels (b) and (c). This figure also illustrates the absence of any zero-energy edge states, thereby confirming our previous statement that the bulk is in a topologically trivial phase. 

The pairs of independent zeroth LLs at opposite valleys are related via an analog of Eq.~\ref{Eq:Mjrn}. In turn, each pair of independent zeroth LLs determine the projections of operators defined for the two sublattices. Upon inclusion of a repulsive Hubbard interaction, as in the case of strain applied along an axis of the BZ, we find that the leading instability is also toward ferromagnetism. A more detailed discussion is given in Sec.~III C and D of the SM. The ferromagnetic state shows a full gap in the bulk. This state would exhibit exponential activation in temperature dependence of the specific heat or the penetration depth to name a few. 

We discuss the case of 3D-T case next. In the absence of strain, we choose a $\Gamma^{-}_{1}$ representation of the tetragonal $D_{4h}$ group~\cite{Sigrist} for the pairing. This leads to two nodal points at $k_{z}= \pm K_{F}$ for a spherical FS, as illustrated in Fig.~\ref{Fig:1}~(b). This type of pairing can be thought of time reversal symmetric analog of the pairing in He-3A \cite{Leggett, Volovik}. As discussed in Sec.~IV A and B of the SM, under uniaxial non-uniform strain along the $x$-direction, and for fixed $k_{z} \approx \pm K_{z}$, the low-energy spectrum of $H_{s=\up}$ along $k_{y}$ consists of single branches of right-moving bulk modes at each valley $(1)$ and $(1')$. For fixed $k_{y} \approx 0$, a flat dispersion emerges along $k_{z}$. Likewise, the opposite-spin sector $H_{s=\dn}$ exhibits two left-moving modes at each of the two valleys. Moreover, we find that for each spin sector, the modes at opposite valleys are related via Eq.~\ref{Eq:Mjrn}. These solutions are consistent with the general discussion of Ref.~\onlinecite{Teo}, which predicts helical Majorana fermions associated with a line defect in a gapped 2D system belonging to the DIII class.

\begin{figure}[t!]
\includegraphics[width=0.75\columnwidth]{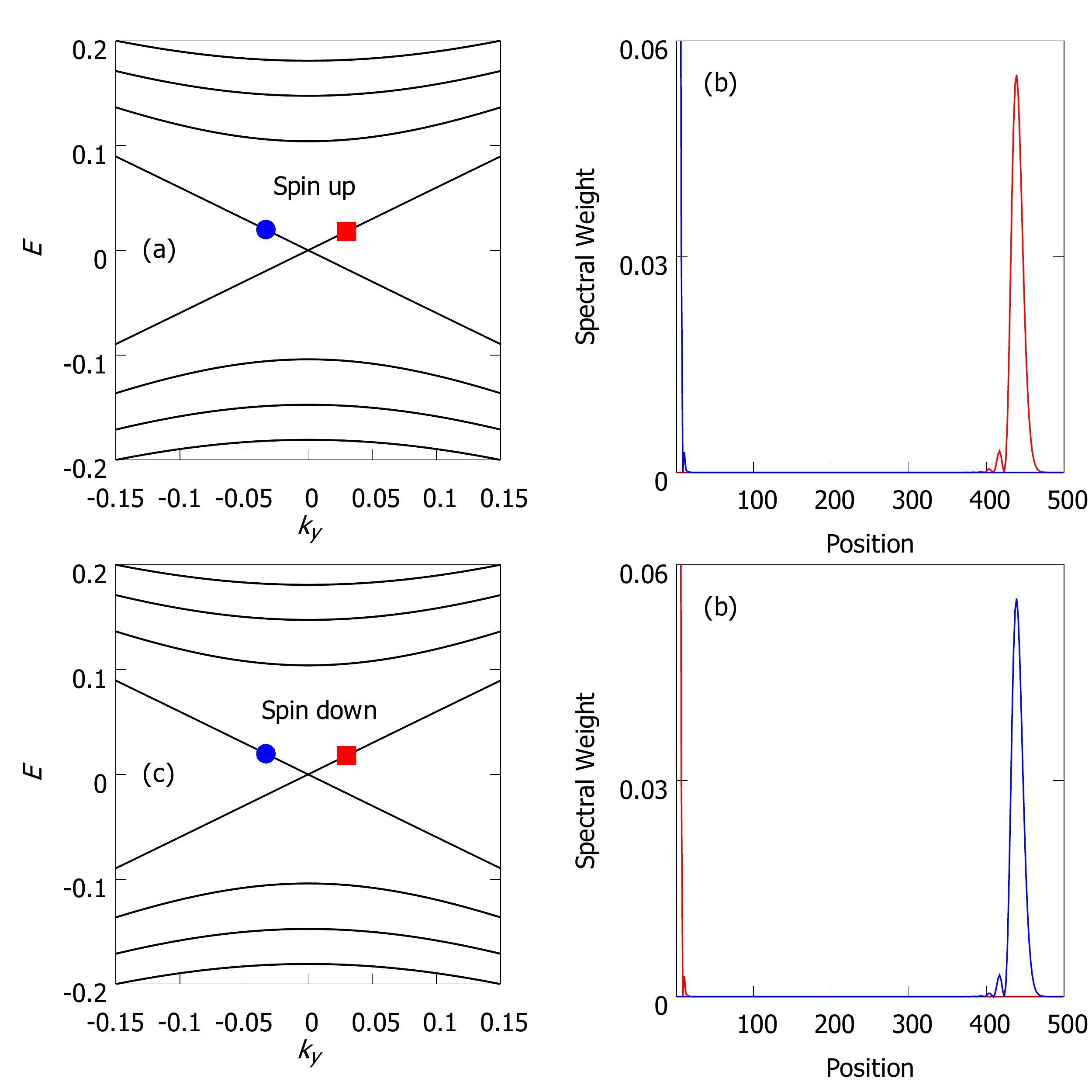}
\caption{3D-T case (a) spectrum of spin up sector under uniaxial strain along $x$ for $k_{z} =K_{F}$ corresponding to valley (1) in Fig.~\ref{Fig:1}~(b), as a function of $k_{y}$. The parameters of the calculation are discussed in the main text.~(b) Spectral weights of the two counter-propagating modes indicated in (a) as functions of position along $x$. Only the right-moving modes are bulk zeroth LL states.~(c) and (d) are the same as (a) and (b) for the spin down sector. Here, the left-moving mode is in the bulk.}
\label{Fig:5}
\end{figure}

The analytical results are confirmed by our lattice calculations. In Fig.~\ref{Fig:5}~(a), we show the spin up spectrum as a function of $k_{y}$ at valley $(1)$ corresponding to $k_{z} \approx K_{F}$. These results were obtained for a chain of 500 sites with a pairing amplitude of $\Delta=10^{-1}$ and strain parameter $\epsilon=9 \times 10^{-3}$. As illustrated by Fig.~\ref{Fig:5}~(b), only one of the two branches corresponds to bulk states, while the other is localized at the edge, confirming the scenario set out in Fig.~\ref{Fig:1}~(c). Furthermore, we see that the two branches are interchanged for the spin-down sector at the same valley, as indicated by Figs.~\ref{Fig:5}~(c) and (d). 

The Fermi fields in the unpaired basis, projected onto the zeroth LL for each spin sector obey a real-space Majorana condition (see SM Sec.~IV C)

\enni \begin{align}
c_{P,\sigma}(\bm{r}) = & e^{iK_{Fz}z} c^{(1)}_{P,\sigma}(\bm{r}) 
+ e^{-iK_{Fz}z} c^{(1')}_{P,\sigma}(\bm{r}) \notag \\
= & (i \sigma_{z})_{\sigma \sigma'} c^{\dag}_{P,\sigma}(\bm{r}),
\label{Eq:Rl_spc_Mjrn}
\end{align}

\enni since the projected operators at each valley are related via $c^{(1')}_{P,\sigma} = \sum_{\sigma'} (i \sigma_{z})_{\sigma \sigma'} c^{(1), \dag}_{P,\sigma'}$. Note that a similar relation does not hold in the singlet case, where zero-modes of \emph{opposite spin} are related to each other. 

In contrast to the 2D-S case, the real-space Majorana nature of the projected fields in the 3D-T case imposes strong constraints on possible residual interactions. It is well-known~\cite{Elliot} that the product of identical real-space Majorana operators reduces to a constant. This reflects the fact that ground-states of Majorana systems do not conserve particle number~\cite{Elliot}. It follows that any local density $c^{\dag}_{\sigma} (\bm{R}_{i}) c_{\sigma}(\bm{R}_{i}) \rightarrow c_{P, \sigma}(\bm{R}_{i}) c_{P,\sigma}(\bm{R}_{i}) = \text{Constant}$. As such, density-density interactions of any range between projected Majorana operators are dynamically trivial. However non-local, four-fermion exchange terms of the form  $U(\bm{R}, \bm{R'}) c_{P,\sigma}(\bm{R}) c_{P,\bar{\sigma}}(\bm{R}) c_{P,\sigma}(\bm{R}')c_{P,\bar{\sigma}}(\bm{R}')$ are allowed. These are unlike spinless Majorana fermions where a minimal interaction is defined on a plaquette \cite{AffleckPRB2015, AffleckPRB2017, AffleckPRB2018, AffleckPRB2019}. To our knowledge, interactions between spinful Majorana fermions have not been proposed in a condensed matter setting. We anticipate that they can lead to more exotic phases which require further investigation. 

\noindent
\emph{Discussion - }
We showed that non-uniform uniaxial strain leads to the formation of spinful Majorana fermions in SCs with a Dirac nodal spectrum. We also demonstrated that the applied strain can induce topological transitions in the bulk of these gapless SCs. The resulting zeroth LLs are localized at the boundary between two topologically distinct phases within the strained sample. In many ways, these bulk states are analogous to the better-known cases of Majorana fermions localized in vortex cores of topological SCs~\cite{Fu}. We note that a similar correspondance between edge states and the zeroth LLs in a strained Weyl semi-metal was pointed out in Ref.~\onlinecite{Grushin}. To our knowledge, the more general mechanism involving a local closing of the gap has not been clearly identified as such.

We estimate a pseudo-magnetic length $l_{B} \approx 410~a$, where $a$ is the inter-atomic distance. We obtain a separation in energy of the LLs $E_{c} \approx 0.11~$meV, corresponding to a characteristic temperature scale of 1.2 K. These estimates were determined using parameters typical of the cuprate class of SCs and are discussed in detail in Sec.~V of the SM.


As we have shown, the direct observation of these gapless states is precluded by various instabilities. In strained 2D d-wave SCs, the flat zeroth-LL band is unstable toward ferromagnetism. One immediate consequence is the coexistence of superconductivity and ferromagnetism in the \emph{bulk} of a strained d-wave SC. The finite magnetic moment in the bulk is screened via the Meissner effect. However, signatures of the resulting screening currents could be detected by SQUID \cite{squid} or NMR experiments. For 3D triplet pairing, the low-energy gapless states are emergent real-space spinful Majorana fermions, which are likely to host more exotic phases.

\emph{Acknowledgements} We thank Joel Moore for fruitful discussions. This work is supported by ASU startup grant.

\bibliography{Interacting_Majoranas_bib_arxiv}

\begin{thebibliography}{30}%
\makeatletter
\providecommand \@ifxundefined [1]{%
 \@ifx{#1\undefined}
}%
\providecommand \@ifnum [1]{%
 \ifnum #1\expandafter \@firstoftwo
 \else \expandafter \@secondoftwo
 \fi
}%
\providecommand \@ifx [1]{%
 \ifx #1\expandafter \@firstoftwo
 \else \expandafter \@secondoftwo
 \fi
}%
\providecommand \natexlab [1]{#1}%
\providecommand \enquote  [1]{``#1''}%
\providecommand \bibnamefont  [1]{#1}%
\providecommand \bibfnamefont [1]{#1}%
\providecommand \citenamefont [1]{#1}%
\providecommand \href@noop [0]{\@secondoftwo}%
\providecommand \href [0]{\begingroup \@sanitize@url \@href}%
\providecommand \@href[1]{\@@startlink{#1}\@@href}%
\providecommand \@@href[1]{\endgroup#1\@@endlink}%
\providecommand \@sanitize@url [0]{\catcode `\\12\catcode `\$12\catcode
  `\&12\catcode `\#12\catcode `\^12\catcode `\_12\catcode `\%12\relax}%
\providecommand \@@startlink[1]{}%
\providecommand \@@endlink[0]{}%
\providecommand \url  [0]{\begingroup\@sanitize@url \@url }%
\providecommand \@url [1]{\endgroup\@href {#1}{\urlprefix }}%
\providecommand \urlprefix  [0]{URL }%
\providecommand \Eprint [0]{\href }%
\providecommand \doibase [0]{https://doi.org/}%
\providecommand \selectlanguage [0]{\@gobble}%
\providecommand \bibinfo  [0]{\@secondoftwo}%
\providecommand \bibfield  [0]{\@secondoftwo}%
\providecommand \translation [1]{[#1]}%
\providecommand \BibitemOpen [0]{}%
\providecommand \bibitemStop [0]{}%
\providecommand \bibitemNoStop [0]{.\EOS\space}%
\providecommand \EOS [0]{\spacefactor3000\relax}%
\providecommand \BibitemShut  [1]{\csname bibitem#1\endcsname}%
\let\auto@bib@innerbib\@empty
\bibitem [{\citenamefont {Guinea}\ \emph {et~al.}(2009)\citenamefont {Guinea},
  \citenamefont {Katsnelson},\ and\ \citenamefont {Geim}}]{Guinea}%
  \BibitemOpen
  \bibfield  {author} {\bibinfo {author} {\bibfnamefont {F.}~\bibnamefont
  {Guinea}}, \bibinfo {author} {\bibfnamefont {M.~I.}\ \bibnamefont
  {Katsnelson}},\ and\ \bibinfo {author} {\bibfnamefont {A.~K.}\ \bibnamefont
  {Geim}},\ }\href@noop {} {\bibfield  {journal} {\bibinfo  {journal} {Nat.
  Phys.}\ }\textbf {\bibinfo {volume} {6}},\ \bibinfo {pages} {30} (\bibinfo
  {year} {2009})}\BibitemShut {NoStop}%
\bibitem [{\citenamefont {Nica}\ and\ \citenamefont {Franz}(2018)}]{Nica}%
  \BibitemOpen
  \bibfield  {author} {\bibinfo {author} {\bibfnamefont {E.~M.}\ \bibnamefont
  {Nica}}\ and\ \bibinfo {author} {\bibfnamefont {M.}~\bibnamefont {Franz}},\
  }\href@noop {} {\bibfield  {journal} {\bibinfo  {journal} {Phys, Rev. B}\
  }\textbf {\bibinfo {volume} {97}},\ \bibinfo {pages} {024520} (\bibinfo
  {year} {2018})}\BibitemShut {NoStop}%
\bibitem [{\citenamefont {Massarelli}\ \emph {et~al.}(2017)\citenamefont
  {Massarelli}, \citenamefont {Wachtel}, \citenamefont {Wei},\ and\
  \citenamefont {Paramekanti}}]{Massarelli}%
  \BibitemOpen
  \bibfield  {author} {\bibinfo {author} {\bibfnamefont {G.}~\bibnamefont
  {Massarelli}}, \bibinfo {author} {\bibfnamefont {G.}~\bibnamefont {Wachtel}},
  \bibinfo {author} {\bibfnamefont {J.~Y.~T.}\ \bibnamefont {Wei}},\ and\
  \bibinfo {author} {\bibfnamefont {A.}~\bibnamefont {Paramekanti}},\
  }\href@noop {} {\bibfield  {journal} {\bibinfo  {journal} {Phys. Rev. B}\
  }\textbf {\bibinfo {volume} {96}},\ \bibinfo {pages} {224516} (\bibinfo
  {year} {2017})}\BibitemShut {NoStop}%
\bibitem [{\citenamefont {Grushin}\ \emph {et~al.}(2016)\citenamefont
  {Grushin}, \citenamefont {W.~F.~Venderbos}, \citenamefont {Vishwanath},\ and\
  \citenamefont {Ilan}}]{Grushin}%
  \BibitemOpen
  \bibfield  {author} {\bibinfo {author} {\bibfnamefont {A.}~\bibnamefont
  {Grushin}}, \bibinfo {author} {\bibfnamefont {J.}~\bibnamefont
  {W.~F.~Venderbos}}, \bibinfo {author} {\bibfnamefont {A.}~\bibnamefont
  {Vishwanath}},\ and\ \bibinfo {author} {\bibfnamefont {R.}~\bibnamefont
  {Ilan}},\ }\href@noop {} {\bibfield  {journal} {\bibinfo  {journal} {Phys.
  Rev. X}\ }\textbf {\bibinfo {volume} {6}},\ \bibinfo {pages} {041046}
  (\bibinfo {year} {2016})}\BibitemShut {NoStop}%
\bibitem [{\citenamefont {Liu}\ \emph {et~al.}(2017)\citenamefont {Liu},
  \citenamefont {Pikulin},\ and\ \citenamefont {Franz}}]{Liu}%
  \BibitemOpen
  \bibfield  {author} {\bibinfo {author} {\bibfnamefont {T.}~\bibnamefont
  {Liu}}, \bibinfo {author} {\bibfnamefont {D.~I.}\ \bibnamefont {Pikulin}},\
  and\ \bibinfo {author} {\bibfnamefont {M.}~\bibnamefont {Franz}},\
  }\href@noop {} {\bibfield  {journal} {\bibinfo  {journal} {Phys. Rev. B}\ ,\
  \bibinfo {pages} {041201(R)}} (\bibinfo {year} {2017})}\BibitemShut {NoStop}%
\bibitem [{\citenamefont {Ghaemi}\ \emph {et~al.}(2012)\citenamefont {Ghaemi},
  \citenamefont {Cayssol}, \citenamefont {Sheng},\ and\ \citenamefont
  {Vishwanath}}]{Ghaemi}%
  \BibitemOpen
  \bibfield  {author} {\bibinfo {author} {\bibfnamefont {P.}~\bibnamefont
  {Ghaemi}}, \bibinfo {author} {\bibfnamefont {J.}~\bibnamefont {Cayssol}},
  \bibinfo {author} {\bibfnamefont {D.~N.}\ \bibnamefont {Sheng}},\ and\
  \bibinfo {author} {\bibfnamefont {A.}~\bibnamefont {Vishwanath}},\
  }\href@noop {} {\bibfield  {journal} {\bibinfo  {journal} {Phys. Rev. Lett.}\
  }\textbf {\bibinfo {volume} {108}},\ \bibinfo {pages} {266801} (\bibinfo
  {year} {2012})}\BibitemShut {NoStop}%
\bibitem [{\citenamefont {Hashimoto}\ \emph {et~al.}(2014)\citenamefont
  {Hashimoto}, \citenamefont {Vishik}, \citenamefont {He}, \citenamefont
  {Devereaux},\ and\ \citenamefont {Shen}}]{Hashimoto}%
  \BibitemOpen
  \bibfield  {author} {\bibinfo {author} {\bibfnamefont {M.}~\bibnamefont
  {Hashimoto}}, \bibinfo {author} {\bibfnamefont {I.~M.}\ \bibnamefont
  {Vishik}}, \bibinfo {author} {\bibfnamefont {R.-H.}\ \bibnamefont {He}},
  \bibinfo {author} {\bibfnamefont {T.~P.}\ \bibnamefont {Devereaux}},\ and\
  \bibinfo {author} {\bibfnamefont {Z.-X.}\ \bibnamefont {Shen}},\ }\href@noop
  {} {\bibfield  {journal} {\bibinfo  {journal} {Nat. Phys.}\ }\textbf
  {\bibinfo {volume} {10}},\ \bibinfo {pages} {483} (\bibinfo {year}
  {2014})}\BibitemShut {NoStop}%
\bibitem [{\citenamefont {Franz}\ and\ \citenamefont
  {Te\u{s}anovi\'{c}}(2000)}]{Franz}%
  \BibitemOpen
  \bibfield  {author} {\bibinfo {author} {\bibfnamefont {M.}~\bibnamefont
  {Franz}}\ and\ \bibinfo {author} {\bibfnamefont {Z.}~\bibnamefont
  {Te\u{s}anovi\'{c}}},\ }\href@noop {} {\bibfield  {journal} {\bibinfo
  {journal} {Phys. Rev. Lett.}\ }\textbf {\bibinfo {volume} {84}},\ \bibinfo
  {pages} {554} (\bibinfo {year} {2000})}\BibitemShut {NoStop}%
\bibitem [{\citenamefont {Sato}\ and\ \citenamefont {Ando}(2017)}]{Sato}%
  \BibitemOpen
  \bibfield  {author} {\bibinfo {author} {\bibfnamefont {M.}~\bibnamefont
  {Sato}}\ and\ \bibinfo {author} {\bibfnamefont {Y.}~\bibnamefont {Ando}},\
  }\href@noop {} {\bibfield  {journal} {\bibinfo  {journal} {Rep. Prog. Phys.}\
  }\textbf {\bibinfo {volume} {80}},\ \bibinfo {pages} {076501} (\bibinfo
  {year} {2017})}\BibitemShut {NoStop}%
\bibitem [{\citenamefont {Teo}\ and\ \citenamefont {Kane}(2010)}]{Teo}%
  \BibitemOpen
  \bibfield  {author} {\bibinfo {author} {\bibfnamefont {J.~C.~Y.}\
  \bibnamefont {Teo}}\ and\ \bibinfo {author} {\bibfnamefont {C.~L.}\
  \bibnamefont {Kane}},\ }\href@noop {} {\bibfield  {journal} {\bibinfo
  {journal} {Phys. Rev. B}\ }\textbf {\bibinfo {volume} {82}},\ \bibinfo
  {pages} {115120} (\bibinfo {year} {2010})}\BibitemShut {NoStop}%
\bibitem [{\citenamefont {de~Gennes}(1999)}]{de_Gennes}%
  \BibitemOpen
  \bibfield  {author} {\bibinfo {author} {\bibfnamefont {P.~G.}\ \bibnamefont
  {de~Gennes}},\ }\href@noop {} {\emph {\bibinfo {title} {Superconductivity of
  Metals and Alloys}}}\ (\bibinfo  {publisher} {Westview, Boulder},\ \bibinfo
  {year} {1999})\BibitemShut {NoStop}%
\bibitem [{\citenamefont {Potter}\ and\ \citenamefont {Lee}(2014)}]{Potter}%
  \BibitemOpen
  \bibfield  {author} {\bibinfo {author} {\bibfnamefont {A.~C.}\ \bibnamefont
  {Potter}}\ and\ \bibinfo {author} {\bibfnamefont {P.~A.}\ \bibnamefont
  {Lee}},\ }\href@noop {} {\bibfield  {journal} {\bibinfo  {journal} {Phys.
  Rev. Lett.}\ }\textbf {\bibinfo {volume} {112}},\ \bibinfo {pages} {117002}
  (\bibinfo {year} {2014})}\BibitemShut {NoStop}%
\bibitem [{\citenamefont {Sigrist}\ and\ \citenamefont {Ueda}(1991)}]{Sigrist}%
  \BibitemOpen
  \bibfield  {author} {\bibinfo {author} {\bibfnamefont {M.}~\bibnamefont
  {Sigrist}}\ and\ \bibinfo {author} {\bibfnamefont {K.}~\bibnamefont {Ueda}},\
  }\href@noop {} {\bibfield  {journal} {\bibinfo  {journal} {Rev. Mod. Phys.}\
  }\textbf {\bibinfo {volume} {63}},\ \bibinfo {pages} {239} (\bibinfo {year}
  {1991})}\BibitemShut {NoStop}%
\bibitem [{\citenamefont {Leggett}(1975)}]{Leggett}%
  \BibitemOpen
  \bibfield  {author} {\bibinfo {author} {\bibfnamefont {A.~J.}\ \bibnamefont
  {Leggett}},\ }\href@noop {} {\bibfield  {journal} {\bibinfo  {journal} {Rev.
  Mod. Phys.}\ }\textbf {\bibinfo {volume} {47}},\ \bibinfo {pages} {331}
  (\bibinfo {year} {1975})}\BibitemShut {NoStop}%
\bibitem [{\citenamefont {Volovik}(2003)}]{Volovik}%
  \BibitemOpen
  \bibfield  {author} {\bibinfo {author} {\bibfnamefont {G.~E.}\ \bibnamefont
  {Volovik}},\ }\href@noop {} {\emph {\bibinfo {title} {The Universe in a
  Helium Droplet}}}\ (\bibinfo  {publisher} {Oxford University Press, NY},\
  \bibinfo {year} {2003})\BibitemShut {NoStop}%
\bibitem [{\citenamefont {Elliott}\ and\ \citenamefont {Franz}(2015)}]{Elliot}%
  \BibitemOpen
  \bibfield  {author} {\bibinfo {author} {\bibfnamefont {S.~R.}\ \bibnamefont
  {Elliott}}\ and\ \bibinfo {author} {\bibfnamefont {M.}~\bibnamefont
  {Franz}},\ }\href@noop {} {\bibfield  {journal} {\bibinfo  {journal} {Rev.
  Mod. Phys.}\ }\textbf {\bibinfo {volume} {87}},\ \bibinfo {pages} {137}
  (\bibinfo {year} {2015})}\BibitemShut {NoStop}%
\bibitem [{\citenamefont {Rahmani}\ \emph {et~al.}(2015)\citenamefont
  {Rahmani}, \citenamefont {Zhu}, \citenamefont {Franz},\ and\ \citenamefont
  {Affleck}}]{AffleckPRB2015}%
  \BibitemOpen
  \bibfield  {author} {\bibinfo {author} {\bibfnamefont {A.}~\bibnamefont
  {Rahmani}}, \bibinfo {author} {\bibfnamefont {X.}~\bibnamefont {Zhu}},
  \bibinfo {author} {\bibfnamefont {M.}~\bibnamefont {Franz}},\ and\ \bibinfo
  {author} {\bibfnamefont {I.}~\bibnamefont {Affleck}},\ }\href@noop {}
  {\bibfield  {journal} {\bibinfo  {journal} {Phys. Rev. B}\ }\textbf {\bibinfo
  {volume} {92}},\ \bibinfo {pages} {235123} (\bibinfo {year}
  {2015})}\BibitemShut {NoStop}%
\bibitem [{\citenamefont {Affleck}\ \emph {et~al.}(2017)\citenamefont
  {Affleck}, \citenamefont {Rahmani},\ and\ \citenamefont
  {Pikulin}}]{AffleckPRB2017}%
  \BibitemOpen
  \bibfield  {author} {\bibinfo {author} {\bibfnamefont {I.}~\bibnamefont
  {Affleck}}, \bibinfo {author} {\bibfnamefont {A.}~\bibnamefont {Rahmani}},\
  and\ \bibinfo {author} {\bibfnamefont {D.}~\bibnamefont {Pikulin}},\
  }\href@noop {} {\bibfield  {journal} {\bibinfo  {journal} {Phys. Rev. B}\
  }\textbf {\bibinfo {volume} {96}},\ \bibinfo {pages} {125121} (\bibinfo
  {year} {2017})}\BibitemShut {NoStop}%
\bibitem [{\citenamefont {Wamer}\ and\ \citenamefont
  {Affleck}(2018)}]{AffleckPRB2018}%
  \BibitemOpen
  \bibfield  {author} {\bibinfo {author} {\bibfnamefont {K.}~\bibnamefont
  {Wamer}}\ and\ \bibinfo {author} {\bibfnamefont {I.}~\bibnamefont
  {Affleck}},\ }\href@noop {} {\bibfield  {journal} {\bibinfo  {journal} {Phys.
  Rev. B}\ }\textbf {\bibinfo {volume} {98}},\ \bibinfo {pages} {245120}
  (\bibinfo {year} {2018})}\BibitemShut {NoStop}%
\bibitem [{\citenamefont {Rahmani}\ \emph {et~al.}(2019)\citenamefont
  {Rahmani}, \citenamefont {Pikulin},\ and\ \citenamefont
  {Affleck}}]{AffleckPRB2019}%
  \BibitemOpen
  \bibfield  {author} {\bibinfo {author} {\bibfnamefont {A.}~\bibnamefont
  {Rahmani}}, \bibinfo {author} {\bibfnamefont {D.}~\bibnamefont {Pikulin}},\
  and\ \bibinfo {author} {\bibfnamefont {I.}~\bibnamefont {Affleck}},\
  }\href@noop {} {\bibfield  {journal} {\bibinfo  {journal} {Phys. Rev. B}\
  }\textbf {\bibinfo {volume} {99}},\ \bibinfo {pages} {085110} (\bibinfo
  {year} {2019})}\BibitemShut {NoStop}%
\bibitem [{\citenamefont {Fu}\ and\ \citenamefont {Kane}(2008)}]{Fu}%
  \BibitemOpen
  \bibfield  {author} {\bibinfo {author} {\bibfnamefont {L.}~\bibnamefont
  {Fu}}\ and\ \bibinfo {author} {\bibfnamefont {C.~L.}\ \bibnamefont {Kane}},\
  }\href@noop {} {\bibfield  {journal} {\bibinfo  {journal} {Phys. Rev. Lett.}\
  }\textbf {\bibinfo {volume} {100}},\ \bibinfo {pages} {096407} (\bibinfo
  {year} {2008})}\BibitemShut {NoStop}%
\bibitem [{\citenamefont {Frolov}\ \emph {et~al.}(2008)\citenamefont {Frolov},
  \citenamefont {Stoutimore}, \citenamefont {Crane}, \citenamefont
  {Van~Harlingen}, \citenamefont {Oboznov}, \citenamefont {Ryazanov},
  \citenamefont {Ruosi}, \citenamefont {Granata},\ and\ \citenamefont
  {Russo}}]{squid}%
  \BibitemOpen
  \bibfield  {author} {\bibinfo {author} {\bibfnamefont {S.}~\bibnamefont
  {Frolov}}, \bibinfo {author} {\bibfnamefont {M.~J.~A.}\ \bibnamefont
  {Stoutimore}}, \bibinfo {author} {\bibfnamefont {T.~A.}\ \bibnamefont
  {Crane}}, \bibinfo {author} {\bibfnamefont {D.~J.}\ \bibnamefont
  {Van~Harlingen}}, \bibinfo {author} {\bibfnamefont {V.~A.}\ \bibnamefont
  {Oboznov}}, \bibinfo {author} {\bibfnamefont {V.~V.}\ \bibnamefont
  {Ryazanov}}, \bibinfo {author} {\bibfnamefont {A.}~\bibnamefont {Ruosi}},
  \bibinfo {author} {\bibfnamefont {C.}~\bibnamefont {Granata}},\ and\ \bibinfo
  {author} {\bibfnamefont {M.}~\bibnamefont {Russo}},\ }\href@noop {}
  {\bibfield  {journal} {\bibinfo  {journal} {Nat. Phys.}\ }\textbf {\bibinfo
  {volume} {4}},\ \bibinfo {pages} {32} (\bibinfo {year} {2008})}\BibitemShut
  {NoStop}%
\bibitem [{\citenamefont {Auerbach}(1994)}]{Auerbach}%
  \BibitemOpen
  \bibfield  {author} {\bibinfo {author} {\bibfnamefont {A.}~\bibnamefont
  {Auerbach}},\ }\href@noop {} {\emph {\bibinfo {title} {Interacting Electrons
  and Quantum Magnetism}}}\ (\bibinfo  {publisher} {Springer-Verlag, N.Y.},\
  \bibinfo {year} {1994})\BibitemShut {NoStop}%
\bibitem [{\citenamefont {Castro}\ \emph {et~al.}(2009)\citenamefont {Castro},
  \citenamefont {Guinea}, \citenamefont {Peres}, \citenamefont {Novoselov},\
  and\ \citenamefont {Geim}}]{Castro}%
  \BibitemOpen
  \bibfield  {author} {\bibinfo {author} {\bibfnamefont {A.~H.}\ \bibnamefont
  {Castro}}, \bibinfo {author} {\bibfnamefont {F.}~\bibnamefont {Guinea}},
  \bibinfo {author} {\bibfnamefont {N.~M.~R.}\ \bibnamefont {Peres}}, \bibinfo
  {author} {\bibfnamefont {K.~S.}\ \bibnamefont {Novoselov}},\ and\ \bibinfo
  {author} {\bibfnamefont {A.~K.}\ \bibnamefont {Geim}},\ }\href@noop {}
  {\bibfield  {journal} {\bibinfo  {journal} {Rev. Mod. Phys.}\ }\textbf
  {\bibinfo {volume} {81}},\ \bibinfo {pages} {109} (\bibinfo {year}
  {2009})}\BibitemShut {NoStop}%
\bibitem [{\citenamefont {Messiah}(1999)}]{Messiah}%
  \BibitemOpen
  \bibfield  {author} {\bibinfo {author} {\bibfnamefont {A.}~\bibnamefont
  {Messiah}},\ }\href@noop {} {\emph {\bibinfo {title} {Quantum Mechanics}}}\
  (\bibinfo  {publisher} {Dover, N. Y.},\ \bibinfo {year} {1999})\ \bibinfo
  {note} {p. 492}\BibitemShut {NoStop}%
\bibitem [{\citenamefont {Jackiw}\ and\ \citenamefont {Rebbi}(1976)}]{Jackiw}%
  \BibitemOpen
  \bibfield  {author} {\bibinfo {author} {\bibfnamefont {R.}~\bibnamefont
  {Jackiw}}\ and\ \bibinfo {author} {\bibfnamefont {C.}~\bibnamefont {Rebbi}},\
  }\href@noop {} {\bibfield  {journal} {\bibinfo  {journal} {Phys. Rev. D}\
  }\textbf {\bibinfo {volume} {13}},\ \bibinfo {pages} {3398} (\bibinfo {year}
  {1976})}\BibitemShut {NoStop}%
\bibitem [{\citenamefont {Balatskii}\ \emph {et~al.}(1986)\citenamefont
  {Balatskii}, \citenamefont {Volovik},\ and\ \citenamefont
  {Konyshev}}]{Balatsky}%
  \BibitemOpen
  \bibfield  {author} {\bibinfo {author} {\bibfnamefont {A.~V.}\ \bibnamefont
  {Balatskii}}, \bibinfo {author} {\bibfnamefont {G.~E.}\ \bibnamefont
  {Volovik}},\ and\ \bibinfo {author} {\bibfnamefont {V.~A.}\ \bibnamefont
  {Konyshev}},\ }\href@noop {} {\bibfield  {journal} {\bibinfo  {journal} {Zh.
  Eksp. Teor. Fiz.}\ }\textbf {\bibinfo {volume} {90}},\ \bibinfo {pages}
  {2038} (\bibinfo {year} {1986})}\BibitemShut {NoStop}%
\bibitem [{\citenamefont {Vozmediano}\ \emph {et~al.}(2010)\citenamefont
  {Vozmediano}, \citenamefont {Katsnelson},\ and\ \citenamefont
  {Guinea}}]{Vozmediano}%
  \BibitemOpen
  \bibfield  {author} {\bibinfo {author} {\bibfnamefont {M.}~\bibnamefont
  {Vozmediano}}, \bibinfo {author} {\bibfnamefont {M.}~\bibnamefont
  {Katsnelson}},\ and\ \bibinfo {author} {\bibfnamefont {F.}~\bibnamefont
  {Guinea}},\ }\href@noop {} {\bibfield  {journal} {\bibinfo  {journal} {Phys.
  Rep.}\ }\textbf {\bibinfo {volume} {496}},\ \bibinfo {pages} {109} (\bibinfo
  {year} {2010})}\BibitemShut {NoStop}%
\bibitem [{\citenamefont {Korshunov}\ \emph {et~al.}(2004)\citenamefont
  {Korshunov}, \citenamefont {Gavrichkov}, \citenamefont {Ovchinnikov},
  \citenamefont {Manske},\ and\ \citenamefont {Eremin}}]{Korshunov}%
  \BibitemOpen
  \bibfield  {author} {\bibinfo {author} {\bibfnamefont {M.~M.}\ \bibnamefont
  {Korshunov}}, \bibinfo {author} {\bibfnamefont {V.~A.}\ \bibnamefont
  {Gavrichkov}}, \bibinfo {author} {\bibfnamefont {S.~G.}\ \bibnamefont
  {Ovchinnikov}}, \bibinfo {author} {\bibfnamefont {D.}~\bibnamefont
  {Manske}},\ and\ \bibinfo {author} {\bibfnamefont {I.}~\bibnamefont
  {Eremin}},\ }\href@noop {} {\bibfield  {journal} {\bibinfo  {journal} {Phys.
  C}\ }\textbf {\bibinfo {volume} {402}},\ \bibinfo {pages} {365} (\bibinfo
  {year} {2004})}\BibitemShut {NoStop}%
\bibitem [{\citenamefont {Mouallem-Bahout}\ \emph {et~al.}(1994)\citenamefont
  {Mouallem-Bahout}, \citenamefont {Gaud\'{e}}, \citenamefont {Calvarin},
  \citenamefont {Gavarri},\ and\ \citenamefont {Carel}}]{Mouallem}%
  \BibitemOpen
  \bibfield  {author} {\bibinfo {author} {\bibfnamefont {M.}~\bibnamefont
  {Mouallem-Bahout}}, \bibinfo {author} {\bibfnamefont {J.}~\bibnamefont
  {Gaud\'{e}}}, \bibinfo {author} {\bibfnamefont {G.}~\bibnamefont {Calvarin}},
  \bibinfo {author} {\bibfnamefont {J.-R.}\ \bibnamefont {Gavarri}},\ and\
  \bibinfo {author} {\bibfnamefont {C.}~\bibnamefont {Carel}},\ }\href@noop {}
  {\bibfield  {journal} {\bibinfo  {journal} {Mater. Lett.}\ }\textbf {\bibinfo
  {volume} {18}},\ \bibinfo {pages} {181} (\bibinfo {year} {1994})}\BibitemShut
  {NoStop}%
\end{thebibliography}%

\pagebreak
\widetext
\begin{center}
\textbf{\large Supplementary Materials for ``Interacting Majorana fermions in strained nodal superconductors''
}
\end{center}

\makeatletter

\renewcommand{\theequation}{S\arabic{equation}}
\renewcommand{\thefigure}{S\arabic{figure}}
\renewcommand{\bibnumfmt}[1]{[S#1]}



The Supplementary Materials contain detailed derivations of our results and discussions and in support of the arguments presented in the main text. Sec.~\ref{Sec:I} reviews important aspects of the pairing Hamiltonian and the Bogoliubov-de Gennes (BdG) equations. It provides general arguments in support of the Majorana nature of the zeroth Landau Levels (LLs) and of the truncation of their Hilbert space, as discussed in the main text. Sec.~\ref{Sec:II} considers the case of 2D d-wave superconductors (SCs) under varying uniaxial strain along one axis of the Brillouin Zone (BZ). We introduce the effective lattice models and present their analytical solutions in the continuum limit. This subsection also describes the gapless solutions which do not correspond to Landau quantization, and briefly elaborates on the connection with the Jackiw-Rebbi equations. We also illustrate the projection of the lattice operators to the zeroth LL sector, and discuss the ensuing ferromagnetic instability. Sec.~\ref{Sec:III} considers the case of 2D d-wave SCs with strain along the direction of a diagonal of the BZ. The major points of Sec.~\ref{Sec:II} are also covered here. Sec.~\ref{Sec:IV} focuses on the case of uniaxial strain in 3D equal-spin triplet superconductors. In particular, it presents the effective lattice models and their analytical solutions in the continuum limit. It also discusses the projection of the lattice operators onto the zeroth LL sector and illustrates their emergent real-space Majorana nature. Sec.\ref{Sec:V} presents our estimates of the pseudo-magnetic length and LL spacing in energy for realistic values of the non-uniform strain in high-$T_{c}$ compounds. Unless stated otherwise, we work in units where $\hbar=1$.
\tableofcontents

\section{General Aspects of the Bogoliubov-de Gennes Hamiltonian}
\label{Sec:I}

In this section, we discuss the general properties of spinful Majorana fermions which occur for time-reversal invariant singlet and equal spin-triplet pairing Hamiltonians under 
applied strain. The general discussion in this section is exemplified in subsequent sections which treat specific cases in detail. 

We define a lattice Hamiltonian in $D=2, 3$ dimensions under applier strain as 

\enni \begin{align}
H_{\text{Pairing}} = & \sum_{\bm{k}} \bm{c}^{\dag}\left(\bm{k} \right) H\left(\bm{k} \right) \bm{c}\left(\bm{k} \right),
\label{Eq:BdG_cnvt}
\end{align}

\enni where spinors are given by

\enni \begin{align}
\bm{c}\left(\bm{k} \right) = & 
\begin{pmatrix}
c_{\up}\left( x_{i}, \bm{k} \right) \\
c_{\dn}\left( x_{i}, \bm{k} \right) \\
c^{\dag}_{\dn}\left( x_{i}, -\bm{k} \right) \\
- c^{\dag}_{\up}\left( x_{i}, -\bm{k} \right) \\
\end{pmatrix}.
\label{Eq:Hmlt_cnvt}
\end{align}

\enni The position $x_{i}, i \in \{1, \hdots, N \}$ indicates the 
position along the direction of the applied uniaxial strain, while 
$\bm{k}$ represents the $D-1$ component conserved momentum along 
the directions which are perpendicular to the applied strain. 
The Hamiltonian can be written as 

\enni \begin{align}
H \left(\bm{k}\right) = & 
\begin{pmatrix}
h \left(x_{i}; x_{j}; \bm{k} \right) \sigma_{0} & \Delta \left( x_{i}; x_{j}; \bm{k} \right) \\
 \Delta^{\dag} \left( x_{i}; x_{j}; \bm{k} \right) & - h^{*}\left( x_{i}; x_{j}; \bm{k} \right) \sigma_{0}
\end{pmatrix}.
\label{Eq:Hmlt}
\end{align}


\enni Due to the presence of the strain, each block is a matrix which 
depends on \emph{both} position indices, as well as the momentum. 

In either singlet or equal-spin triplet pairing cases, we 
can separate Hamiltonian into two independent spin sectors as

\enni \begin{align}
H(\bm{k}) =
\begin{pmatrix}
H_{s= \up}(\bm{k}) & 0 \\
0 & H_{s=\dn}(\bm{k})
\end{pmatrix},
\end{align}

\enni corresponding to spin $(\up \dn)$, $(\dn \up)$ sectors in the singlet case, and spin $(\up \up)$ and $(\dn \dn)$ sectors in the equal-spin triplet cases. 

The solutions of each $H_{s}$ are then obtained 
via a standard Bogoliubov-de Gennes (BdG) ansatz~\cite{Elliot}

\enni \begin{align}
\gamma_{s, n}(\bm{k}) = & \sum_{x_{i}, \sigma} u^{*}_{s\sigma, n}(x_{i}, \bm{k}) c_{\sigma}\left(x_{i}, \bm{k} \right) + v^{*}_{s\sigma, n}(x_{i}, \bm{k}) c^{\dag}_{\sigma} \left(x_{i},-\bm{k} \right),
\label{Eq:Anst_sngl}
\end{align} 

\enni for energies $E \ge 0$. We likewise define the inverse transformation as

\enni \begin{align}
c_{\sigma}\left(x_{i}, \bm{k}\right)= \sum_{n, s} u_{s\sigma, n}(x_{i}, \bm{k}) \gamma_{s, n}(k) + 
v^{*}_{s \sigma, n}(x_{i}, -\bm{k}) \gamma^{\dagger}_{s, n}(-\bm{k}).
\label{Eq:Invr_BdG_trns}
\end{align}

\enni The index $s$ also labels the  conserved spin of the BdG quasiparticles. 
For the case of singlet pairing $u_{s\sigma} = \delta_{s\sigma} u_{s}$ and 
$v_{s\sigma}= (i\sigma_{y})_{s\sigma} v_{s}$. In the case of equal-spin triplet pairing $u$ has an identical form while $v_{s\sigma}= - (\sigma_{y})_{s\sigma} v_{s}$.
The index $n$ represents all other labels. 
In the vicinity of the nodal points, it labels the bulk Landau Levels (LLs). 

As discussed below and in the following sections, we find that only half of the degrees-of-freedom for the zeroth LL sector are independent since $\gamma_{s, 0}(\bm{k}) = M_{ss'} \gamma^{\dag}_{s', 0}(-\bm{k})$, where $M= \sigma_{y}$ for 2D-S, and $M= -i \sigma_{z}$ for singlet and triplet cases, respectively. Note that the inverse BdG transformation in Eq.~\ref{Eq:Invr_BdG_trns} implicitly assumes that the two components are independent of each other. Therefore, in order to preserve a well-defined BdG transformation, we restrict the $u, v$ indices to half of the allowed momenta. This procedure is illustrated in Sec.~\ref{Sec:IId}.

The Majorana nature of the zeroth LLs follows from the general p-h symmetry of the Hamiltonian. Indeed, each $H_{s}$ sector obeys the BdG eqs. 

\enni \begin{align}
H_{s}\left(\bm{k}\right) \Phi_{s,n}(\bm{k}) = & E_{n}(\bm{k}) \Phi_{s, n}(\bm{k}), 
\end{align}

\enni where the BdG spinor is

\enni \begin{align}
\Phi_{s, n}(\bm{k}) = & 
\begin{pmatrix}
u_{s\sigma, n}\left(x_{i}, \bm{k} \right) \\
v_{s\sigma', n}\left(x_{i}, \bm{k} \right) \\
\end{pmatrix}.
\end{align}

We define a particle-hole (p-h) transformation

\enni \begin{align}
C_{\text{Singlet}}= 
(-i \tau_{y}) \bm{1}_{N\times N} K,
\label{Eq:Sngl_ph_trns}
\end{align}

\enni for singlet and 

\enni \begin{align}
C_{\text{Triplet}}= 
\tau_{x} \bm{1}_{N\times N} K,
\end{align}

\enni for triplet pairing, respectively. The $N \times N$ identity matrix is defined for the position indices and $K$ represents complex conjugation. The two different forms are due to the distinct parities of the two types of pairing. The p-h 
transformation acts on the Hamiltonians for each of the two spin sectors as~\cite{Sato}

\enni \begin{align}
C H_{s}\left(\bm{k}\right) C^{-1} = - H_{s}\left(-\bm{k}\right).
\end{align}

\enni As is well-know, this implies that the eigenstates of  $H_{s}(\bm{k})$ come in pairs of opposite energy:

\enni \begin{align}
H_{s}\left(\bm{k}\right) \left[C \Phi_{s, n}(-\bm{k})\right] = & -E_{n}(-\bm{k}) \left[ C \Phi_{s,n}(-\bm{k}) \right]. 
\end{align}

\enni When strain induces a \emph{single, zeroth} LL mode, p-h symmetry imposes an additional constraint. Since only 
one of either $\Phi_{s, 0}(\bm{k})$ or $C \Phi_{s, 0}(-\bm{k})$ is a non-trivial solution of $H_{\sigma \sigma'}(\bm{k})$, we require

\enni \begin{align}
C \Phi_{s, 0}(-\bm{k}) = &
\begin{pmatrix}
 \pm v^{*}_{s, 0}(x_{i}, -\bm{k}) \\
u^{*}_{s, 0}(x_{i}, -\bm{k})
\end{pmatrix} \\
= & 0.
\end{align}

\enni This necessarily implies that the Hilbert space of the zeroth LLs must be truncated to half of all allowed momenta $\bm{k}$. 

The above argument excludes the case where the spinor is invariant 
under the p-h transformation: $C \Psi_{s}(-\bm{k}) \propto \Psi_{s}(\bm{k})$. This is confirmed by explicit solutions presented in Sec.~\ref{Sec:IIc}.

Finally, we discuss the case where zero-energy LLs occur in pairs, as discussed in detail in Sec.~\ref{Sec:IIIb} and~\ref{Sec:IIIc}. Consider the case where $H_{s}$ are invariant under a mirror plane involving the conserved momentum $k$: 

\enni \begin{align}
H_{s}(k) = H_{s}(-k).
\end{align}

\enni Together with the p-h transformation

\enni \begin{align}
-H_{s}\left(-k\right) \left[C \Phi_{s, n}(k)\right] = & E_{n}(\bm{k}) \left[ C \Phi_{s,n}(\bm{k}) \right],
\end{align}

\enni this implies

\enni \begin{align}
H_{s}\left(k\right) \left[C \Phi_{s, n}(k)\right] = & -E_{n}(\bm{k}) \left[ C \Phi_{s,n}(\bm{k}) \right].
\label{Eq:Mrr_smmt}
\end{align}

\enni Therefore, zero-energy solutions occur in pairs and are determined by $\Psi_{s, 0}$ and $C \Psi_{s, 0}(k)$.

\section{Two dimensional d-wave pairing with strain along one nodal axis}
\label{Sec:II}

In this section, we discuss the case of a 2D $d_{xy}$ SC with strain along one of the nodal axes of the pairing. In Sec.~\ref{Sec:IIa}, we discuss the effective 1D lattice Hamiltonians which are appropriate under applied uniaxial strain. An effective low-energy Hamiltonian is derived in Sec.~\ref{Sec:IIb} and it's detailed LL solutions are presented in Sec.~\ref{Sec:IIc}. Sec.~\ref{Sec:IId} discuss the resulting projection of the lattice operators onto the zeroth LL sector. The last subsection is devoted to the discussion of a short-range Hubbard interaction projected onto the zeroth LLs and the ensuing ferromagnetic instability. 

\subsection{1D Lattice Hamiltonian in the presence of uniaxial strain}
\label{Sec:IIa}

Using the convention of Eq.~\ref{Eq:Hmlt_cnvt}, we consider a simple lattice model for a 2D d-wave consisting of 
nearest-neighbor (NN) hopping together with next-nearest-neighbor (NNN) $d_{xy}$ pairing on a square lattice:

\enni \begin{align}
H= & H_{TB} + H_{\Delta} \\
H_{TB} = & \sum_{x_{i}, y_{i}, j, \sigma} t(x_{i}) \left[c^{\dag}_{\sigma}(x_{i}, y_{i})c_{\sigma}(x_{i}+\delta_{x,j}, y_{i}+\delta_{y, j})+ c^{\dag}_{\sigma}(x_{i}+\delta_{x,j}, y_{i}+\delta_{y, j}) c_{\sigma}(x_{i}, y_{i}) \right] - \mu c^{\dag}_{\sigma}(x_{i}, y_{i})c_{\sigma}(x_{i}, y_{i})\\
H_{\text{Pair}} = & \sum_{x_{i}, y_{i}, j, \sigma, \sigma'} (i \sigma_{y})_{\sigma \sigma'} \Delta(\bm{\delta'}_{j}) \left[
c_{\sigma}(x_{i}, y_{i})c_{\sigma'}(x_{i}+\delta'_{x,j}, y_{i}+\delta'_{y, j}) + c_{\sigma}(x_{i}, y_{i})c_{\sigma'}(x_{i}-\delta'_{x,j}, y_{i}-\delta'_{y, j}) \right]+ \text{H.c.}.
\label{Eq:d_wv_hmlt}
\end{align}

\enni Note that the pairing amplitude is independent of spin indices $\sigma, \sigma'$. We define the NN and NNN vectors $\bm{\delta}_{1}=(a, 0), \bm{\delta}_{2}=(0, a)$ and $\bm{\delta'}_{1}=(a, a), \bm{\delta'}_{2}=(-a, a)$, where $a$ is the NN distance. The sign change of the $d_{xy}$ pairing is induced via $\Delta(\bm{\delta'}_{1/2})=\pm \Delta$. Without any strain, we have $t(x_{i})=t$ and the resulting dispersion shows gapless excitations around the four momenta $(\pm K_{F}, 0)$ and $(0, \pm K_{F})$, which denote the positions of four valleys denoted by $(2), (2')$ and $(1), (1')$, respectively. 

We allow for the effects of strain by introducing a position dependent hopping along the $x$ direction

\enni \begin{align}
t(x_{i}) = & t + \delta t(x_{i}) \\
\delta t(x_{i}) = & t \epsilon x_{i},
\end{align}

\enni where $\epsilon$ is approximately proportional to the gradient of the strain along $x$ (Sec.~\ref{Sec:Va}). The modification to the unperturbed hopping $t$ is due to the 
non-uniform stretching of the NN bonds. For a microscopic derivation of this term, we refer the reader to the supplementary material of Ref.~\onlinecite{Nica}. We can in principle allow an analogous  variation in the pairing amplitudes. However, as shown in the supplementary material of Ref.~\onlinecite{Nica}, in the low-energy limit, such a contribution 
can be eliminated via a gauge transformation and does not change our results qualitatively. 

Our central assumption is that the contribution to the hopping due to 
strain is small compared to any of the parameters of the unperturbed system over the finite length of the sample i.e. $\text{max}(\delta t(x_{i})) \ll \Delta, t$. As a consequence, we assume that the applied strain does not destroy the d-wave superconductor. 

We assume periodic boundary conditions along the $y$ direction and apply a Fourier transform:

\enni \begin{align}
c_{\sigma}(x_{i}, y_{i}) = & \frac{1}{N_{y}} \sum_{k_{y}} e^{ik_{y} y_{i}} c_{\sigma}(x_{i}, k_{y}),
\end{align}

\enni where the 1D Brillouin Zone (BZ) is defined by $k_{y} \in [-\pi/a, \pi/a]$. The Hamiltonian becomes 

\enni \begin{align}
H_{TB} = & \sum_{i, k_{y}, \sigma} t(x_{i}) \left[c^{\dag}_{\sigma}(x_{i}, k_{y})c_{\sigma}(x_{i}+a, k_{y})+ c^{\dag}_{\sigma}(x_{i}+a, k_{y}) c_{\sigma}(x_{i}, k_{y}) \right] - \mu c^{\dag}_{\sigma}(x_{i}, k_{y})c_{\sigma}(x_{i}, k_{y}) \notag \\
+ & 2t(x_{i}) \cos \left(k_{y}a \right) c^{\dag}_{\sigma}(x_{i}, k_{y})c_{\sigma}(x_{i}, k_{y}) \\
H_{\text{Pair}} = & \sum_{x_{i}, k_{y}, \sigma, \sigma'} (- i \Delta)(i\sigma_{y})_{\sigma \sigma'} \sin\left(k_{y}a \right)\left[ c_{\sigma}\left(x_{i}, -k_{y} \right) c_{\sigma'}\left(x_{i} - a, k_{y} \right) - c_{\sigma}\left(x_{i}, -k_{y} \right) c_{\sigma'}\left(x_{i} + a, k_{y} \right)   \right] + \text{H.c.}
\end{align}

\enni Using the ansatz for singlet pairing in Eq.~\ref{Eq:Anst_sngl}, we obtain the BdG Eqs for each $H_{s}$ sector as

\enni \begin{align}
 & t(x_{i}) \left[u_{s, n}(x_{i+1}, k) + u_{s, n}(x_{i-1}, k) \right] -\mu u_{s, n}(x_{i}, k) + 2t(x_{i}) \cos \left(ka \right) u_{s, n}(x_{i}, k) \notag \\  
 + & i \Delta \sin\left(ka\right)  \left[ v_{s, n}(x_{i-1}, k) - v_{s, n}(x_{i+1}, k) \right] 
 =  E_{n}(k) u_{s, n}(x_{i}, k) \notag \\
 & \notag 
\\
& -i \Delta \sin\left(ka\right)  \left[ u_{s, n}(x_{i+1}, k) - u_{s, n}(x_{i-1}, k) \right] \notag \\
- & \bigg\{ 
 t(x_{i}) \left[v_{s, n}(x_{i+1}, k) + v_{s, n}(x_{i-1}, k) \right] -\mu v_{s, n}(x_{i}, k) + 2t(x_{i}) \cos \left(k_{y}a \right) v_{s, n}(x_{i}, k) \bigg\} 
 = E_{n}(k) v_{s\sigma, n}(x_{i}, k),
 \label{Eq:BdG_sngl}
\end{align}

\enni where we dropped the $y$ index.

\subsection{Low-energy effective Hamiltonian under applied uniaxial strain}
\label{Sec:IIb}

We consider the low-energy limit of the lattice BdG eqs. 
As we are interested in solutions in the bulk of a sample, 
we consider an infinite system and ignore the effects of the boundary 
along the direction of the strain. Our numerical results justify 
this approximation. 
 
Note that the lattice and continuum fields are related via the Wannier states~\cite{Auerbach}

\enni \begin{align}
c_{\sigma}\left(x_{i}, y_{i} \right) = &\int dx dy \phi^{*}(x, y; x_{i}, y_{i}) \Psi_{\sigma}(x, y) \\
\Psi_{\sigma}(x, y) = & \sum_{x_{i}, y_{j}} \phi(x, y; x_{i}, y_{j}) c_{\sigma}(x_{i}, y_{i}).
\end{align}

\enni A partial Fourier transform along the $y$ direction gives

\enni \begin{align}
c_{\sigma}\left(x_{i}, k \right) = &\int dx \frac{dk'}{2\pi} \phi^{*}(x, k'; x_{i}, k) \Psi_{\sigma}(x, k').
\end{align}

\enni For sufficiently large systems, we allow the Fourier transform of the Wannier function to be sharply peaked 

\enni \begin{align}
\phi^{*}(x, k'; x_{i}, k) \approx 2\pi \delta(k-k') \tilde{\phi}(x; x_{i}).
\end{align}

\enni Using these expression in Eq.~\ref{Eq:Anst_sngl}, we obtain

\enni \begin{align}
\gamma_{s, n}(k) = & \sum_{\sigma} \int dx \left[ u^{*}_{s\sigma, n}(x, k) \Psi_{\sigma}\left(x, k \right) - \text{sgn}(\bar{\sigma}) v^{*}_{s\sigma, n}(x, k) \Psi^{\dag}_{\bar{\sigma}} \left(x,-k \right) \right].
\end{align}

\enni where we defined the continuum version of the BdG coefficients as

\enni \begin{align}
u^{(\alpha),*}_{\sigma, n}(x, k) = & \sum_{x_{i}}  \tilde{\phi^{*}}(x; x_{i}) u^{(\alpha),*}_{\sigma, n} (x_{i}, k)  \\
v^{(\alpha),*}_{\sigma, n}(x, k) = & \sum_{x_{i}}  \tilde{\phi}(x; x_{i}) v^{(\alpha),*}_{\sigma, n} (x_{i}, k)
\end{align}


\enni We assume for simplicity that the Wannier states can be chosen to be real. We can formally switch to the continuum by multiplying 
the each one of eqs.~\ref{Eq:BdG_sngl} by the Wannier state $\tilde{\phi}$ and summing over all $i$. We subsequently drop the index $i$.
 
In the low-energy limit and for momenta in the vicinity of 
each valley s.t. $k \approx K^{(\alpha)}_{Fy} + q_{y}$, we employ the approximation 

\enni \begin{align}
u_{s, n}(x, K^{(\alpha)}_{Fy} + q) \approx e^{iK^{(\alpha)}_{Fx} x}  u^{(\alpha)}_{s, n}(x, q) \\
v_{s, n}(x, K^{(\alpha)}_{Fy} + q)  \approx e^{iK^{(\alpha)}_{Fx} x}  v^{(\alpha)}_{s, n}(x, q),
\end{align}

\enni where $\alpha \in \{1,1',2,2'\}$ is a valley index and 
$(u, v)^{(\alpha)}$ are envelope functions which are assumed to vary slowly on the scale of the lattice. 
Such an approximation is well-known in the context of graphene~\cite{Castro}. We introduce a cutoff scale $|q| \le \Lambda$.

Our ansatz allows for a separation of the 
solutions in terms of valley index. We simplify common 
phase factors $e^{iK_{x} x}$ and 
expand the envelope functions s.t. 

\enni \begin{align}
e^{iK^{(\alpha)}_{Fx} a} u^{(\alpha)}_{s, n}(x+a, q) + e^{-iK^{(\alpha)}_{Fx} a} u^{(\alpha)}_{s, n}(x-a, q) 
\approx 2 \cos\left(K^{(\alpha)}_{Fx}a \right) u^{(\alpha)}_{s, n}(x, q) + 2a i \sin\left(K^{(\alpha)}_{Fx}a \right) \partial_{x} 
u^{(\alpha)}_{s, n}(x, q) \\
e^{-iK_{Fx}a} v^{(\alpha)}_{s, n}(x-a, q) - e^{iK_{Fx}a} v^{(\alpha)}_{s, n}(x+a, q) \approx 
- 2i \sin\left(K^{(\alpha)}_{Fx}a \right) v^{(\alpha)}_{s, n}(x, q) 
-2 a \cos\left( K^{(\alpha)}_{Fx}a \right) \partial_{x} v^{(\alpha)}_{s, n}(x, q).  
\end{align}

\enni We can further expand

\enni \begin{align}
\cos \left(ka \right) \approx \cos \left(K^{(\alpha)}_{Fy}a \right) 
- \sin \left(K^{(\alpha)}_{Fy}a \right) qa \\
\sin \left(ka \right) \approx \sin \left(K^{(\alpha)}_{Fy}a \right) 
+ \cos \left(K^{(\alpha)}_{Fy}a \right) qa.
\end{align}

We keep only terms in Eqs.~\ref{Eq:BdG_sngl} which are first order 
in \emph{either of the three small quantities} $\delta t(x), \partial_{x}, q$. We note that the zeroth order terms vanish either due to presence of a Fermi surface or to the vanishing of the pairing at each valley. To first order, the first Eq. of \ref{Eq:BdG_sngl} becomes

\enni \begin{align}
\left[ 2\delta t(x) \cos\left(K^{(\alpha)}_{Fx}a \right) + 2ta \sin\left(K^{(\alpha)}_{Fx}a \right) i \partial_{x} -2ta \sin\left(K^{(\alpha)}_{Fy}a \right) q + 2\delta t(x) \cos\left(K^{(\alpha)}_{Fy}a \right)  \right] u^{(\alpha)}_{s, n}(x, q) \notag \\
 \left[ -2 \Delta a \sin\left(K^{(\alpha)}_{Fy}a \right) \cos\left(K^{(\alpha)}_{Fx}a \right) i \partial_{x} 
 + 2 \Delta a \cos\left(K^{(\alpha)}_{Fy}a \right) \sin\left(K^{(\alpha)}_{Fx}a \right) q \right] v^{(\alpha)}_{s, n}(x, q) = E_{n}(k) u^{(\alpha)}_{s, n}(x_{i}, k).
\end{align}

\enni We define 

\enni \begin{align}
v_{F} = & 2ta \sin\left(K_{F}a \right) \\
v_{\Delta} = & 2\Delta a \sin\left(K_{F}a \right) \label{Eq:Sngl_Frmi_vlct} \\
A(x) = & 2 \delta t(x) \left[ \cos\left(K_{Fx}a  \right) 
+ \cos\left(K_{Fy}a \right) \right],
\end{align}

\enni where $K_{F} = \text{max}(|K^{(\alpha)}_{Fx}|, |K^{(\alpha)}_{Fy}|)$ is independent of $\alpha$ and positive. The BdG Eqs can be recast as

\enni \begin{align}
v_{F} \left[ \text{sgn}\left(K^{(\alpha)}_{Fx} \right) i \partial_{x} - \text{sgn}\left(K^{(\alpha)}_{Fy} \right) q + \frac{A}{v_{F}}\right] u^{(\alpha)}_{s, n}(x_{i}, q) 
+ v_{\Delta}\left[ - \text{sgn}\left(K^{(\alpha)}_{Fy} \right) i \partial_{x} + 
\text{sgn}\left(K^{(\alpha)}_{Fx} \right) q   \right]v^{(\alpha)}_{s, n}(x_{i}, q) \notag  \\
=  E_{n}(q)\left(K^{(\alpha)}_{Fy} + q \right) u^{(\alpha)}_{s, n}(x, q) \notag \\
v_{\Delta}\left[ - \text{sgn}\left(K^{(\alpha)}_{Fy} \right) i \partial_{x} + 
\text{sgn}\left(K^{(\alpha)}_{Fx} \right) q   \right]u^{(\alpha)}_{s, n}(x, q) - v_{F} \left[ \text{sgn}\left(K^{(\alpha)}_{Fx} \right) i \partial_{x} - \text{sgn}\left(K^{(\alpha)}_{Fy} \right) q + \frac{A}{v_{F}}\right] v^{(\alpha)}_{s, n}(x, q) \notag \\
= E_{n}(q)\left(K^{(\alpha)}_{Fy} + q \right) v^{(\alpha)}_{s, n}(x, q).
\label{Eq:Bdg_sngl_effc}
\end{align}

\subsection{Landau levels from the low-energy Hamiltonian}
\label{Sec:IIc}

We first focus on the (1), (1') valleys which 
correspond to $\bm{K_{F}}=(0, \pm K_{F})$. The strain-induced 
vector potential induces LLs about the nodal momenta. By contrast, LLs do not emerge in the vicinity of valleys (2), (2'), as shown further down below. 

The BdG Eqs. in the vicinity of valley (1) are given by 

\enni \begin{align}
v_{F} \left( -  q + \frac{A}{v_{F}}\right) u^{(1)}_{s, n}(x, q) 
+ v_{\Delta}\left( - i \partial_{x} \right)v^{(1)}_{s, n}(x, q) = & E_{n}(K^{(1)}_{Fy} + q) u^{(1)}_{s, n}(x, q) \notag \\
v_{\Delta}\left( - i \partial_{x} \right)u^{(1)}_{s, n}(x, q) - v_{F} \left( -  q + \frac{A}{v_{F}}\right) v^{(1)}_{s, n}(x, q)
= & E_{n}(K^{(1)}_{Fy} + q) v^{(1)}_{s, n}(x, q).
\label{Eq:BdG_sngl_vll_1}
\end{align}
 
\enni We apply a unitary transformation  

\enni \begin{align}
U= \frac{1}{\sqrt{2}}\left( \sigma_{0} + i \sigma_{x} \right),
\label{Eq:Trns_sngl}
\end{align}

\enni where $\sigma_{x}$ acts on the space of the BdG spinor. 
The new Bdg Eqs. for valley (1) read

\enni \begin{align}
\begin{pmatrix}
0 & v_{\Delta}\left( - i \partial_{x} \right) -i v_{F} \left( -  q + \frac{e \tilde{A}}{v_{F}}\right) \\
v_{\Delta}\left( - i \partial_{x} \right) + i v_{F} \left( -  q + \frac{e\tilde{A}}{v_{F}}\right) & 0
\end{pmatrix}
\begin{pmatrix}
\tilde{u}^{(1)}_{s, n}(x, q) \\
\tilde{v}^{(1)}_{s, n}(x, q)
\end{pmatrix}
= & E_{n}(q)
\begin{pmatrix}
\tilde{u}^{(1)}_{s, n}(x, q) \\
\tilde{v}^{(1)}_{s, n}(x, q)
\end{pmatrix},
\label{Eq:Jckw_Rbbi}
\end{align}

\enni where we introduced the effective vector potential 

\enni \begin{align}
\bm{\tilde{A}} = \left(0, \frac{A(x)}{e} \right).
\end{align}

\enni Recalling that $\tilde{A}(x) = Bx$ \emph{by construction},
 we define a pseudo-magnetic field as

\enni \begin{align}
B = \bm{\nabla} \times \bm{A}.
\end{align}

\enni We now clearly see that the BdG equations correspond 
to Dirac fermions in the presence of a uniform magnetic field in the Landau gauge~\cite{Castro}. The well-known solutions for $n \neq 0$ are given by the Landau levels

\enni \begin{align}
\begin{pmatrix}
\tilde{u}^{(1)}_{s, n}(x_{i}, q) \\
\tilde{v}^{(1)}_{s, n}(x_{i}, q)
\end{pmatrix}
= & \frac{1}{\sqrt{2}} 
\begin{pmatrix}
 \psi_{n-1}(x, q) \\
\pm \psi_{n}(x, q)
\end{pmatrix},
\end{align}

\enni with energies 

\enni \begin{align}
E_{n} = & \pm \omega_{c} \sqrt{n}. 
\end{align}

\enni The cyclotron frequencies and magnetic lengths are 
defined by 

\enni \begin{align}
\omega_{c} = & \sqrt{2} \frac{v_{\Delta}}{l_{B}} \\
l_{B} = & \sqrt{\frac{v_{\Delta}}{e B}}.
\end{align}

\enni $\Psi_{n}(x, q)$ represent harmonic oscillator 
wavefunctions~\cite{Messiah}

\enni \begin{align}
\psi_{n}(\xi) = & \sqrt{\frac{1}{l_{B}}} u_{n}(\xi), \\
u_{n} =  & \frac{1}{\pi^{1/4} \sqrt{2^{n} n!}} e^{-\xi^{2}/2} H_{n}(\xi), \\
\sum_{n} u_{n}(\xi) u_{n}(\xi') = & \delta(\xi-\xi'), \label{Eq:Hrmt_cmpl}\\
\int^{\infty}_{-\infty} d\xi u_{n}(\xi) u_{m}(\xi) = & \delta_{nm}.
\end{align}

\enni where $H_{n}$ are Hermite polynomials and the dimensionless 
variable is $\xi = (x/l_{B}) - (v_{F}/v_{\Delta})l_{B}q$. 
The $\psi_{n}$ functions obey the normalization condition

\enni \begin{align}
\int dx \psi_{n}(x, q) \psi_{m}(x,q) = &  \int \frac{dx}{\lambda l_{B}} u_{n}(\xi) u_{m}(\xi) \notag \\
= & \int d\xi  u_{n}(\xi) u_{m}(\xi) \notag \\
= & \delta_{nm}.
\end{align}

Referring to the BdG Eqs. (\ref{Eq:Bdg_sngl_effc}), 
for valley (1') we have 

\enni \begin{align}
v_{F} \left( q + \frac{A}{v_{F}}\right) u^{(1')}_{s, n}(x, q) 
+ v_{\Delta}\left( i \partial_{x} \right)v^{(1')}_{s, n}(x, q) = & E_{n}(q) u^{(1')}_{s, n}(x, q) \notag \\
v_{\Delta}\left( i \partial_{x} \right)u^{(1')}_{s, n}(x, q) - v_{F} \left( q + \frac{A}{v_{F}}\right) v^{(1')}_{s, n}(x, q)
= & E_{n}(q) v^{(1')}_{s, n}(x, q).
\end{align}

\enni These can be obtained from Eqs.~\ref{Eq:BdG_sngl_vll_1} for valley (1) via complex conjugation and sending $q \rightarrow -q$. 
This procedure amounts to a time-reversal operation and illustrates the that this symmetry is preserved by the applied strain. 

In the un-rotated basis, the solutions for the 
non-zeroth LLs are 

\enni \begin{align}
\begin{pmatrix}
u^{(\alpha)}_{s, n}(x, q) \\
v^{(\alpha)}_{s, n}(x, q)
\end{pmatrix}
= & \frac{1}{2} 
\begin{pmatrix}
 \psi_{n-1}(x, \text{sgn}(\alpha) q) - i \text{sgn}(\alpha) \psi_{n}(x, \text{sgn}(\alpha) q) \\
-i \text{sgn}(\alpha) \psi_{n-1}(x, \text{sgn}(\alpha) q) + \psi_{n}(x, \text{sgn}(\alpha) q)
\end{pmatrix},
\end{align}

\enni where $\text{sgn}(\alpha)= \pm 1$ for $\alpha \in \{1, 1'\}$.

The \emph{zeroth} LL in the original basis is determined by 

\enni \begin{align}
\begin{pmatrix}
u^{(\alpha)}_{s, 0}(x_{i}, q) \\
v^{(\alpha)}_{s, 0}(x_{i}, q)
\end{pmatrix} 
= & \frac{\psi_{0}(x, \text{sgn}(\alpha) q)}{\sqrt{2}}
\begin{pmatrix}
 -i \text{sgn}(\alpha) \\
 1 
\end{pmatrix}.
\end{align}



The continuum version of the BdG ansatz in Eq.~\ref{Eq:Anst_sngl} is given by 

\enni \begin{align}
\gamma^{(\alpha)}_{s, n}(q) = & \int dx \left[u^{(\alpha),*}_{s\sigma, n}(x, q) \Psi^{(\alpha)}_{\sigma}\left(x, q \right) +  v^{(\alpha),*}_{s\sigma, n}(x, q) \Psi^{(\bar{\alpha}),\dag}_{\sigma} \left(x,-q \right)\right].
\end{align} 

\enni In the spin-singlet case $u_{s\sigma}= \delta_{s\sigma}$ and similarly for $v_{s\sigma}= (i\sigma_{y})_{s \sigma} v_{s}$. The explicit expressions are

\enni \begin{align}
\gamma^{(1)}_{\up,0}(q) = & \int dx \frac{  \psi_{0}(x, q)}{\sqrt{2}} \left[ i \Psi^{(1)}_{\up}\left(x, q \right) + \Psi^{(1'),\dag}_{\dn}\left(x, - q \right) \right] \label{Eq:Sngl_0_LL_1}\\
\gamma^{(1)}_{\dn,0}(q) =  & \int dx \frac{ \psi_{0}(x, q)}{\sqrt{2}} \left[ i \Psi^{(1)}_{\dn}\left(x, q \right) - \Psi^{(1'), \dag}_{\up}\left(x_{i}, - q \right) \right] \\
\gamma^{(1')}_{\up,0}(q) = & \int dx \frac{ \psi_{0}(x, -q)}{\sqrt{2}} \left[ -i \Psi^{(1')}_{\up}\left(x, q \right) +\Psi^{(1),\dag}_{\dn}\left(x, - q \right) \right] \\
\gamma^{(1')}_{\dn,0}(q) = & \int dx \frac{ \psi_{0}(x, -q)}{\sqrt{2}} \left[ -i \Psi^{(1')}_{\dn}\left(x, q \right) -  \Psi^{(1), \dag}_{\up}\left(x, - q \right) \right]. \label{Eq:Sngl_0_LL_2}
\end{align}

\enni It follows that 

\enni \begin{align}
\gamma^{(1')}_{\up,0}(q) = & i \gamma^{(1), \dag}_{\dn,0}(-q) \label{Eq:Mjrn_sngl_1}\\
\gamma^{(1')}_{\dn,0}(q) = & -i \gamma^{(1), \dag}_{\up,0}(-q) \label{Eq:Mjrn_sngl_2}.
\end{align}

We now comment on the solutions at valleys (2), (2'). 
Based on Eqs.~\ref{Eq:Bdg_sngl_effc}, the first of these is subject to the following eqs.

\enni \begin{align}
v_{F} \left(  i \partial_{x} + \frac{A}{v_{F}}\right) u^{(2)}_{s, n}(x, q) 
+ v_{\Delta} q v^{(2)}_{\sigma, n}(x, q) 
=  E_{n}\left(q \right) u^{(2)}_{s, n}(x, q) \notag \\
v_{\Delta} 
 q u^{(\alpha)}_{s, n}(x, q) - v_{F} \left(  i \partial_{x} + \frac{A}{v_{F}}\right) v^{(2)}_{s, n}(x, q) = E_{n}\left( q \right) v^{(2)}_{s, n}(x, q).
\end{align}

\enni These do not result in the emergence of LLs. The vector 
potential can be taken into account by incorporating a suitable global phase:

\enni \begin{align}
u^{(2)}_{s, n}(x, q) = e^{i \int^{x}_{x_{0}} dx' \frac{A(x')}{v_{F}} } \tilde{u}^{(2)}_{s, n}(x, q) \\
u^{(2)}_{s, n}(x, q) = e^{i \int^{x}_{x_{0}} dx' \frac{A(x')}{v_{F}} } \tilde{u}^{(2)}_{s, n}(x, q).
\end{align}

\enni The solutions correspond to linearly-dispersing, counter-propagating modes centered around $K_{Fy}=0$. 

We finally comment on the topological nature of the \emph{zeroth} LLs. To illustrate, we consider the zero-energy solutions of Eqs.~\ref{Eq:Jckw_Rbbi}. By canceling factors of $i$ these eqs. can be reduced to 

\enni \begin{align}
\left[ v_{\Delta}(-i \sigma_{y}) \partial_{x} + m(x) \sigma_{x} \right] \Phi^{(1)}_{0}(x,q) = 0, 
\end{align}

\enni where $m(x)= v_{F}q - A(x)$. These are identical to the  well-known Jackiw-Rebbi eqs.~\cite{Jackiw}. The 
latter host a topologically-protected domain-wall solution centered about the point where $m(x)$ changes sign. Going beyond the low-energy approximation, the zeroth LL modes are expected to persist as topologically-protected modes centered around line-defects, in accordance with the findings of Ref.~\onlinecite{Teo}.

\subsection{Projection of lattice operators onto the zeroth LL modes}
\label{Sec:IId}

We consider solutions deep in the bulk of a strained 
sample and thus ignore contributions from any edge states. As before, 
we can pass onto the continuum via an expansion using the Wannier states. The continuum version of the inverse BdG transformation is 

\enni \begin{align} 
\Psi^{(\alpha)}_{\sigma}\left(x, q \right)= \sum_{n, s} u^{(\alpha)}_{s\sigma, n}(x, q) \gamma^{(\alpha)}_{s, n}(k) +
v^{(\bar{\alpha}),*}_{s \sigma, n}(x, -q) \gamma^{(\bar{\alpha}),\dagger}_{s, n}(-q).
\label{Eq:Sngl_invr_BdG}
\end{align} 

\enni Note that for the zeroth LL contribution, only two out of the four states in Eqs.~\ref{Eq:Sngl_0_LL_1}-\ref{Eq:Sngl_0_LL_2} are independent. This necessarily implies that the inverse BdG transformation must be restricted to half of the the zeroth LL states. Without loss of generality, we write the expansion of the fields as 

\enni \begin{align}
\Psi^{(\alpha)}_{\sigma}(x, q) =  \Psi^{(\alpha)}_{P,\sigma}
+ \frac{1}{2} \sum_{|n| \ge 1} \left[ \psi_{n-1}(x, \text{sgn}(\alpha)q) - i \text{sgn}(\alpha) \psi_{n}(x, \text{sgn}(\alpha) q) \right] \gamma^{(\alpha)}_{\sigma, n}(q) \notag \\ 
- \text{sgn}(\sigma) \left[i \text{sgn}(\bar{\alpha}) \psi_{n-1}(x, -\text{sgn}(\bar{\alpha}) q) + \psi_{n}(x, -\text{sgn}(\bar{\alpha}) q) \right] \gamma^{(\bar{\alpha}), \dag}_{\bar{\sigma}, n}(-q) ,
\label{Eq:Cmpl_expn}
\end{align} 

\enni where we introduced the projection onto the zeroth LL

\enni \begin{align}
\Psi^{(1)}_{P,\up}(x,q) = & \frac{ -i\psi_{0}(x, q)}{\sqrt{2}} \gamma^{(1)}_{\up, 0}(q) \\
\Psi^{(1)}_{P,\dn}(x,q) = & \frac{ -i\psi_{0}(x, q)}{\sqrt{2}} \gamma^{(1)}_{\dn, 0}(q) \\
\Psi^{(1')}_{P,\up}(x, q) = & \frac{-\psi_{0}(x, -q)}{\sqrt{2}} \gamma^{(1), \dag}_{\dn, 0}(-q) \\
\Psi^{(1')}_{P,\dn}(x, q) = & \frac{\psi_{0}(x, -q)}{\sqrt{2}} \gamma^{(1), \dag}_{\up, 0}(-q).
\end{align}

\enni We chose to keep both positive and negative values of $q$, but  retained only those contributions from valley $(1)$. These expressions are consistent with Eqs.~\ref{Eq:Sngl_0_LL_1}-\ref{Eq:Sngl_0_LL_2}. Using the completeness of the Hermite polynomials (Eq.~\ref{Eq:Hrmt_cmpl}) it is straightforward to 
verify that the fields in Eq.~\ref{Eq:Cmpl_expn} satisfy canonical anti-commutation relations. 

Via Eqs.~\ref{Eq:Mjrn_sngl_1}-\ref{Eq:Mjrn_sngl_2}, and \ref{Eq:Sngl_invr_BdG}, one can verify that, without any restriction on the Hilbert space of the zeroth LLs, the inverse BdG transformation is indeed singular. The projections onto the zeroth LL are themselves not independent since 

\enni \begin{align}
\Psi^{(1')}_{P,\sigma}(x, q) = & i \text{sgn}(\sigma) \Psi^{(1), \dagger}_{P,\bar{\sigma}}(x, -q).
\label{Eq:Sngl_prjc_Mjrn}
\end{align}

\enni We now illustrate that Eq.~\ref{Eq:Sngl_prjc_Mjrn} is independent of the choice of Hilbert space for the zeroth LL. 
Consider retaining both valley indices, but restricting the momenta $q$ to positive values. The contribution of the zeroth LLs is given by

\enni \begin{align}
\Psi^{(1)}_{P,\up}(x,q) = & \frac{\psi_{0}(x, q)}{\sqrt{2}} \left[ -i \theta(q) \gamma^{(1)}_{\up, 0}(q) - \theta(-q)\gamma^{(1'), \dag}_{\dn, 0}(-q)  \right] \\
\Psi^{(1)}_{P,\dn}(x,q) = & \frac{ \psi_{0}(x, q)}{\sqrt{2}} \left[ -i \theta(q) \gamma^{(1)}_{\dn, 0}(q) + \theta(-q)\gamma^{(1'), \dag}_{\up, 0}(-q) \right] \\
\Psi^{(1')}_{P,\up}(x, q) = & \frac{\psi_{0}(x, -q)}{\sqrt{2}}\left[ i \theta(-q) \gamma^{(1')}_{\up, 0}(q) -\theta(q) \gamma^{(1), \dag}_{\dn, 0}(-q) \right] \\
\Psi^{(1')}_{P,\dn}(x, q) = &  \frac{\psi_{0}(x, -q)}{\sqrt{2}}\left[ i \theta(-q) \gamma^{(1')}_{\dn, 0}(q) + \theta(q) \gamma^{(1), \dag}_{\up, 0}(-q) \right].
\end{align}

\enni Using Eqs.~\ref{Eq:Mjrn_sngl_1}-\ref{Eq:Mjrn_sngl_2} one can verify that these satisfy the relations in Eqs.~\ref{Eq:Sngl_prjc_Mjrn}.

The Fourier transforms of the projected operators, defined by

\enni \begin{align}
\Psi^{(\alpha)}_{P, \sigma}\left(x, y \right) = & \int^{\Lambda}_{-\Lambda} \frac{dq}{2\pi} e^{iqy} \Psi^{(\alpha)}_{P, \sigma}\left(x, q \right), 
\label{Eq:Rl_spc_expn}
\end{align}

\enni likewise obey

\enni \begin{align}
\Psi^{(\bar{\alpha})}_{P, \sigma}\left(x, y \right) = & i \text{sgn}(\sigma) \Psi^{(\alpha), \dag}_{P, \bar{\sigma}}\left(x, y \right).
\label{Eq:Sngl_rl_spc_Mjrn}
\end{align}

\enni In addition, they are subject to the anti-commutation relations

\enni \begin{align}
\{ \Psi^{(\alpha)}_{\sigma}, \Psi^{(\alpha'), \dagger}_{\sigma'} \} = & -i \text{sgn}(\sigma') \{\Psi^{(\alpha)}_{\sigma}, \Psi^{(\bar{\alpha'})}_{\bar{\sigma'}} \} \notag \\
 = & \delta_{\sigma, \sigma'}  \frac{1}{2}\int^{\Lambda}_{-\Lambda} \left( \frac{dq}{2\pi} \right)  e^{iq(y-y')} \bigg\{  \psi_{0}(x, q) \psi_{0}(x', q) \bigg\} \notag \\
= & \delta_{\sigma, \sigma'} \frac{1}{2 l_{B} \sqrt{\pi}} \int^{\Lambda}_{-\Lambda} \left( \frac{dq}{2\pi} \right)  e^{iq(y-y')} e^{-\left( \frac{x}{l_{B}} - l_{B} \lambda q \right)^{2}/2} e^{-\left( \frac{x'}{l_{B}} - l_{B}\lambda q \right)^{2}/2} \notag \\
= & \delta_{\sigma, \sigma'}  \frac{1}{2 l_{B} \sqrt{\pi}} 
\int^{\Lambda}_{-\Lambda} \left( \frac{dq}{2\pi} \right) e^{-l^{2}_{B} \lambda^{2} q^{2} + \lambda q(x+x') + iq(y-y')} e^{- \frac{x^{2}}{2  l_{B}^{2}}} 
e^{- \frac{x'^{2}}{2 l_{B}^{2}}} \notag \\
\approx & \delta_{\sigma, \sigma'} \frac{1}{2\lambda l_{B}\sqrt{\pi}} \sqrt{\frac{\pi}{l^{2}_{B}}} \frac{1}{2\pi} e^{\frac{\left[ (x+x') + i(y-y') \right]^{2}}{4l^{2}_{B} \lambda^{2}}}
e^{-\frac{x^{2}}{2 l^{2}_{B}}} e^{-\frac{x'^{2}}{2 l^{2}_{B}}} \notag \\
\approx & \delta_{\sigma, \sigma'}\frac{1}{4 \pi \lambda l^{2}_{B}} e^{-\frac{( x - x')^{2}}{4 \lambda^{2} l^{2}_{B}}} e^{-\frac{(y - y')^{2}}{4 \lambda^{2} l^{2}_{B}}} e^{\frac{i (x+x')(y-y')}{2 \lambda l^{2}_{B}}} 
\label{Eq:Sngl_cmmt}
\end{align}

\enni \begin{align}
\{ \Psi^{(\alpha)}_{\sigma}, \Psi^{(\alpha)}_{\sigma'} \} =0.
\end{align}

\enni where $\lambda=v_{F}/v_{\Delta}$. In the fourth line we approximated the integral by a Gaussian in the limit $\Lambda l_{B} \rightarrow \infty$. This approximation is justified for weak pseudo-magnetic fields considered here. We identify the coefficient as half of the density of LLs 

\enni \begin{align}
n_{0} = \frac{1}{2 \pi \lambda l^{2}_{B}},
\end{align}

\enni by using the well-known formula for the degeneracy of LLs 

\enni \begin{align}
N_{LL} = \frac{L_{x}L_{y}}{2\pi l^{2}_{B}},
\end{align}

\enni where $L_{x}, L_{y}$ are the lengths of the system in either direction and we accounted for the anisotropy between the Fermi velocities via the factor $\lambda$.

The field operator projected onto the zeroth LLs gives 

\enni \begin{align}
\Psi_{P, \sigma}(x, y) = & e^{iK_{F}y} \Psi^{(1)}_{P, \sigma}\left(x, y \right) 
+  e^{-iK_{F}y} 
\Psi^{(1')}_{P,\sigma} \left(x, y \right).
\label{Eq:Sngl_prjc}
\end{align}

\subsection{Ferromagnetism at mean-field level}
\label{Sec:IIe}

In order to investigate many body instabilities of the zeroth LL, we consider a repulsive, short-range Hubbard interaction in the continuum limit

\enni \begin{align}
H_{U} = U \int d^{2}r \Psi^{\dag}_{\up}(\bm{r})\Psi_{\up}(\bm{r})\Psi^{\dag}_{\dn}(\bm{r})\Psi_{\dn}(\bm{r}).
\end{align}

\enni Using Eq.~\ref{Eq:Sngl_prjc}, we project this term onto 
the zeroth LLs:

\enni \begin{align}
H_{U,P} = U \int d^{2}r \left[ \rho^{(1)}_{P, \uparrow} \rho^{(1)}_{P, \downarrow} + \rho^{(1)}_{P, \uparrow} \rho^{(1')}_{P, \downarrow} + \rho^{(1')}_{P, \uparrow} \rho^{(1)}_{P, \downarrow} 
+ \rho^{(1')}_{P, \uparrow} \rho^{(1')}_{P, \downarrow} +  \Psi^{(1), \dagger}_{P, \uparrow} \Psi^{(1')}_{P, \uparrow} \Psi^{(1'), \dagger}_{P, \downarrow} \Psi^{(1)}_{P, \downarrow} +  \Psi^{(1'), \dagger}_{P, \uparrow} \Psi^{(1)}_{P, \uparrow} \Psi^{(1), \dagger}_{P, \downarrow} \Psi^{(1')}_{P, \downarrow} \right].
\label{Eq:sngl_vll_hmlt}
\end{align}

\enni Note that we have only kept terms which lack rapidly oscillating phase factors such as $e^{\pm 2iK_{F}x}$. 
We anticipate that in the long-wavelength limit considered here, the  terms we kept are the most relevant. We have also ignored any contributions from boundary terms. 

Using Eq.~\ref{Eq:Sngl_rl_spc_Mjrn}, we can formally eliminate the degrees of freedom at valley $(1')$ in favor of their analogoues at valley $(1)$:

\enni \begin{align}
\rho^{(1)}_{P\uparrow} \rho^{(1')}_{P\downarrow} \rightarrow \Psi^{(1), \dagger}_{P\uparrow} \Psi^{(1)}_{P\uparrow} \Psi^{(1)}_{P\uparrow} \Psi^{(1), \dagger}_{P\uparrow} \\
\rho^{(1')}_{P\uparrow} \rho^{(1)}_{P\downarrow} \rightarrow \Psi^{(1)}_{P\downarrow} \Psi^{(1), \dagger}_{P\downarrow} \Psi^{(1), \dag}_{P\dn} \Psi^{(1)}_{P\dn} \\
\rho^{(1')}_{P\uparrow} \rho^{(1')}_{P\downarrow} \rightarrow 
\Psi^{(1)}_{P\downarrow} \Psi^{(1), \dagger}_{P\downarrow} \Psi^{(1)}_{P\uparrow} \Psi^{(1), \dagger}_{P\uparrow} \\
\Psi^{(1), \dagger}_{\uparrow} \Psi^{(1')}_{\uparrow} \Psi^{(1'), \dagger}_{\downarrow} \Psi^{(1)}_{\downarrow} 
\rightarrow   - \Psi^{(1), \dagger}_{P\uparrow}  \Psi^{(1), \dagger}_{P\downarrow}  \Psi^{(1)}_{P\uparrow} \Psi^{(1)}_{P\downarrow} \\
\Psi^{(1'), \dagger}_{P\uparrow} \Psi^{(1)}_{P\uparrow} \Psi^{(1), \dagger}_{P\downarrow} \Psi^{(1')}_{P\downarrow} \rightarrow - \Psi^{(1)}_{P\downarrow} \Psi^{(1)}_{P\uparrow}  \Psi^{(1), \dagger}_{P\downarrow} \Psi^{(1), \dagger}_{P\uparrow}.
\end{align}

We take advantage of the fact that the anti-commutators of the projected operators is finite and proportional to half of the density of zeroth LLs per spin (Eq.~\ref{Eq:Sngl_cmmt}). This way the Hamiltonian reduces to 

\enni \begin{align}
H_{U,P} =& U \int d^{2}r \left[ 4 \rho^{(1}_{P \uparrow} \rho^{(1)}_{P\downarrow}  - n_{0} (\rho^{(1)}_{P\uparrow} + \rho^{(1)}_{P\downarrow}) + \frac{1}{2}n^{2}_{0} \right]. \\
\end{align}

\enni We write

\enni \begin{align}
\rho^{(1)}_{\uparrow} = & \frac{1}{2}\left[ 2S^{(1)}_{z} + \rho^{(1)} \right] \\
\rho^{(1)}_{\dn} = & -\frac{1}{2}\left[ 2S^{(1)}_{z} - \rho^{(1)} \right],
\end{align}

\enni where 

\enni \begin{align}
S^{(1)}_{z} = & \frac{1}{2} \left( \rho^{(1)}_{\up} - \rho^{(1)}_{\dn}  \right)\\
\rho^{(1)} = & \rho^{(1)}_{\up} + \rho^{(1)}_{\dn}.
\end{align}

\enni The Hamiltonian can be re-written as

\enni \begin{align}
H_{U,P} =& U \int d^{2}r \left[ - 4 \left( S^{(1)}_{P, z} \right)^{2}  
+ \left( \rho^{(1)}_{P} - \frac{n_{0}}{2} \right)^{2} -\frac{n^{2}_{0}}{2} \right]
\end{align}

\enni The Hamiltonian can be decoupled at mean-field level in the ferromagnetic channel. For repulsive interactions, the pairing instability is energetically unfavorable. A possible competing spin-density order is not allowed, since we have eliminated the degrees of freedom at valley $(1')$. 

Taking the local spin-density and total density as variational parameters, we take the expectation value of $H_{U,P}$. From the expression above, it is clear that a half-filled ground state corresponding to $\rho^{(1)}_{P} = n_{0}/2$ is energetically favored.  

Using \ref{Eq:Rl_spc_expn}, the spin-part is given by

\enni \begin{align}
\braket{H_{U, P, \text{Spin}}} = &  - \left( \frac{\sqrt{2\pi } U}{\lambda l_{B}} \right) \int \left( \frac{dk}{2 \pi} \right) \left( \frac{dq}{2 \pi} \right) \left( \frac{dQ}{2 \pi} \right) e^{ -l^{2}_{B} \left\{ \frac{\left[ Q+(k-q) \right]^{2} + Q^{2} }{2} \right\}} \braket{\gamma^{(1), \dagger}_{k+Q \uparrow} \gamma^{(1)}_{k\uparrow} - \gamma^{(1), \dagger}_{k+Q \downarrow} \gamma^{(1)}_{k\downarrow}} 
\braket{\gamma^{(1), \dagger}_{q-Q \uparrow} \gamma^{(1)}_{q\uparrow} - \gamma^{(1), \dagger}_{q-Q \downarrow} \gamma^{(1)}_{q\downarrow}}.
\end{align}

\enni The Gaussian term ensures that the dominant contributions are 
due to momenta $k, q, Q$ which are within all within a scale set by $1/l_{B}$. At half-filling, it is clear that this term is maximally negative at ferromagnetic alignment. 

In conclusion, a ferromagnetic ground-state is favored at mean-field level. The resulting Stoner spectrum consists of flat bands at 

\enni \begin{align}
E_{q} \approx &  \left( \frac{\sqrt{2\pi} U }{\lambda l_{B}} \right) \frac{1}{(2\pi)^{2}}  \int^{\infty}_{-\infty} dk  e^{ -l^{2}_{B} \frac{(k-q)^{2}}{2}} \notag \\
= & \pm  \frac{n_{0}U}{2}.
\end{align}

\section{Two dimensional d-wave pairing with strain off the nodal axes}
\label{Sec:III}

In this section we discuss a 2D d-wave SC with strain applied off the nodal axes of the pairing. In Sec.~\ref{Sec:IIIa}, we introduce the effective 1D lattice Hamiltonians under uniaxial strain. We also discuss their enhanced symmetry and sublattice structure, which are important differences w.r.t. the models of Sec.~\ref{Sec:IIa}. In Sec.~\ref{Sec:IIIb}, we introduce the low-energy, continuum limit of the BdG eqs. and obtain their solutions. We find two degenerate zeroth LLs in the vicinity of each of the two valleys with BdG coefficients which are subject to the constraint imposed by the combined mirror and p-h symmetry discussed in the main text and below. Sec.~\ref{Sec:IIIc} presents a comparison of the analytical solutions with the numerical results. Sec.~\ref{Sec:IIId} is devoted to a discussion of the projection of the lattice operators onto the zeroth LL section, while Sec.~\ref{Sec:IIIe} illustrates the ensuing ferromagnetic instability.

\subsection{1D Lattice Hamiltonian}
\label{Sec:IIIa}

\begin{figure}[ht!]
\includegraphics[width=0.75\columnwidth]{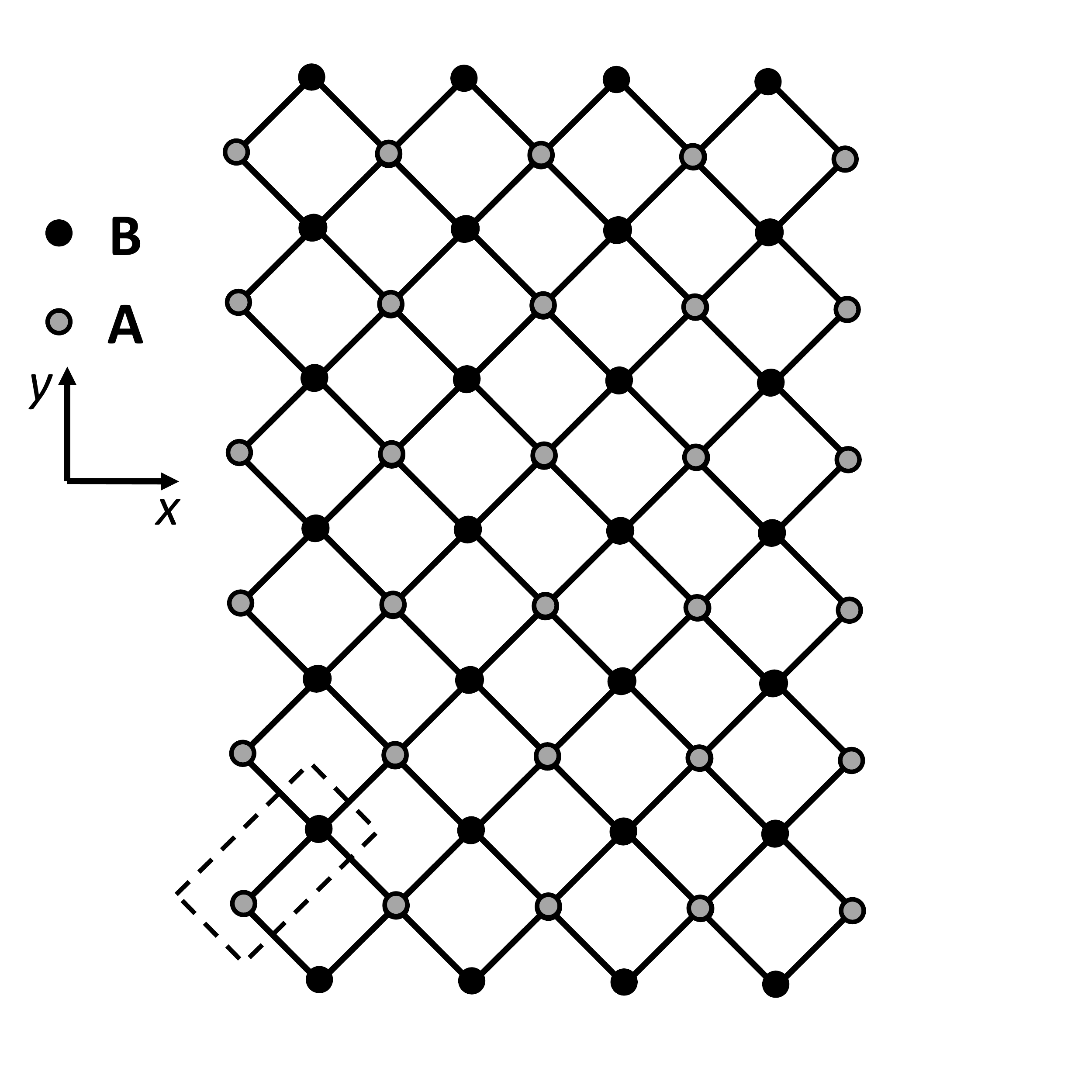}
\caption{Lattice model for 2D-S with strain along the diagonal of the BZ. The $x, y$ directions are rotated wrt the axes of the BZ. In order to preserve translational symmetry in the $y$ direction along the edges, we use a two-site unit cell, illustrated by the dashed line. The strain is along the $x$ direction.}
\label{Fig:S2}
\end{figure}

We consider the tight-binding Hamiltonian in Eq.~\ref{Eq:d_wv_hmlt} without strain and rotate by $\pi/4$. In order to preserve translational symmetry along the rotated $y$ direction at the edges
we choose a two-site unit cell. This lattice is illustrated in Fig.~\ref{Fig:S2}.

\enni We subsequently introduce 
a spatially-varying tight-binding parameter $t(x_{i})$, where $x$ is along the diagonal of the un-rotated system. Applying a Fourier transform along the $y$ direction we obtain

\enni \begin{align}
H = & H_{TB} + H_{\Delta} \\
H_{TB}= & \sum_{i=1, \sigma}^{N_{x}} 2t(x_{i}) 
\cos\left(\frac{k_{y}a}{\sqrt{2}} \right) 
c^{\dagger}_{A, \sigma}(x_{i}, k_{y}) c_{B, \sigma}(x_{i}, k_{y}) +
2t\left(x_{i}+\frac{a}{\sqrt{2}} \right) 
\cos\left(\frac{k_{y}a}{\sqrt{2}} \right) 
c^{\dagger}_{B, \sigma}(x_{i}, k_{y}) c_{A, \sigma}(x_{i+1}, k_{y}) +
\text{H.c.} \notag \\
- & \mu c^{\dagger}_{A, \sigma}(x_{i}, k_{y}) c_{A, \sigma}(x_{i}, k_{y}) 
- \mu c^{\dagger}_{B, \sigma}(x_{i}, k_{y}) c_{B, \sigma}(x_{i}, k_{y})\notag \\
H_{\Delta} = & \sum_{i=1}^{N_{x}} \sum_{\sigma \sigma'} 
(i \sigma_{y})_{\sigma \sigma'} \Delta \big[ c^{\dag}_{\sigma}(A, x_{i}, k_{y}) c^{\dag}_{A, \sigma'}(x_{i+1},-k_{y}) 
+ c^{\dag}_{A, \sigma}(x_{i+1}, k_{y}) c^{\dag}_{A, \sigma'}(x_{i},-k_{y}) \notag \\
- & 2 \cos\left(k_{y} a \sqrt{2} \right)
c^{\dagger}_{A, \sigma}(x_{i}, k_{y}) c_{A, \sigma'}(x_{i},-k_{y}) \big] + (A \leftrightarrow B) + \text{H.c.} 
\end{align}

\enni The position index runs over all even values while the momentum is along a folded 1D BZ. Without strain, the effective chain model is periodic by translation via $a \sqrt{2}$ or $i \rightarrow i+1$. Subsequently, we distinguish between two inequivalent sublattices which we label by $A, B$.

We consider the ansatz in Eq.~\ref{Eq:Anst_sngl} modified to take into account the sublattice structure:

\enni \begin{align}
\gamma_{s, n}(k) = & \sum_{x_{i}, \sigma} \bigg\{ u^{*}_{s\sigma, A, n}(x_{i}, k) c_{A,\sigma}\left(x_{i}, k \right) + v^{*}_{s\sigma, A, n}(x_{i}, k) c^{\dag}_{A, \sigma} \left(x_{i},-k \right) \notag \\
+ & u^{*}_{s\sigma, B, n}(x_{i}, k) c_{B, \sigma}\left(x_{i}, k \right) + v^{*}_{s\sigma, B, n}(x_{i}, k) c^{\dag}_{B, \sigma} \left(x_{i},-k \right) \bigg\}.
\end{align} 

\enni Using the convention in Eq.~\ref{Eq:BdG_cnvt} where $u_{s\sigma}=\delta_{s\sigma} u_{s}$ and $v_{s\sigma}= (i\sigma_{y})_{s\sigma} v_{s}$, the singlet pairing Hamiltonian becomes independent of the spin indices $\sigma$, which we subsequently drop for simplicity. 

The BdG eqs. for sublattice B are given by 

\begin{align}
 2 &
\cos \left( \frac{k_{y}a}{\sqrt{2}} \right) 
\left[ t(x_{i}) u_{s, A, n}(x_{i}, k_{y}) + t(x_{i}+ a/\sqrt{2}) u_{s, A,n}(x_{i+1}, k_{y}) \right] - 
\mu u_{s, B, n}(x_{i}, k_{y}) \notag \\
+ & \Delta \left[ v_{s, B, n}(x_{i+1}, k_{y})
+  v_{s, B, n}(x_{i-1}, k_{y}) \right] 
-  2 \Delta \cos\left(k_{y} a \sqrt{2} \right)v_{s, B, n}(x_{i}, k_{y}) =  E_{n}( k_{y}) u_{s, B, n}(x_{i}, k_{y}) \label{Eq:Dgnl_lttc_BdG_1}\\
\notag \\
& \Delta \left[u_{s,B,n}(x_{i+1}, k_{y}) 
+  u_{s,B,n}(x_{i-1}, k_{y}) \right] 
-  2 \Delta \cos\left(k_{y} a \sqrt{2} \right)u_{s, B, n}(x_{i}, k_{y}) \notag \\
-&  \left\{ 2 
\cos\left(\frac{k_{y}a}{\sqrt{2}} \right) 
\left[ t(x_{i}) v_{s, A, n}(x_{i}, k_{y}) + t(x_{i}+a/\sqrt{2}) v_{s, A, n}(x_{i+1}, k_{y}) \right]  
- \mu v_{s,B,n}(x_{i}, k_{y}) \right\}
=  E_{n} (k_{y}) v_{s,B,n}(x_{i}, k_{y}).
\label{Eq:Dgnl_lttc_BdG_2}
\end{align}

\enni The corresponding eqs. for sublattice $A$ can be obtained by simply exchanging $A, B$ indices. 

Finally, we note that the BdG equations for each sublattice are invariant under $k \rightarrow -k$, as discussed in the main text. 

\subsection{Landau levels in the continuum limit}
\label{Sec:IIIb}

In the limit of zero strain, the bonding solution of Eqs.~\ref{Eq:Dgnl_lttc_BdG_1}-\ref{Eq:Dgnl_lttc_BdG_2} exhibits nodes at four momenta $(\pm K_{F}, \pm K_{F})$, which we label by valley indices $(1), (2), (1'), (2')$ in a clockwise direction. 
The anti-bonding solution is generically gapped and will be ignored in the following. 

We extend the ansatz for the pristine case to the strained system 
by writing the wavefunctions as

\enni \begin{align}
u_{\sigma, A, n}(x_{i})(K_{Fy} + q_{y}) = & e^{iK^{(\alpha)}_{Fx} x_{i}} u^{(\alpha)}_{\sigma, A, n}(x_{i}, q_{y}) +
e^{iK^{\bar{(\alpha)}}_{Fx} x_{i}} u^{(\bar{\alpha})}_{\sigma, A, n}(x_{i}, q_{y}) \\
u_{\sigma, B, n}(x_{i})(K_{Fy} + q_{y}) = & e^{iK^{(\alpha)}_{Fx}  ( x_{i} + a/\sqrt{2})} u^{(\alpha)}_{\sigma, A, n}(x_{i} + a/\sqrt{2}, q_{y}) +
e^{iK^{\bar{(\alpha)}}_{Fx} (x_{i} + a/\sqrt{2})} u^{(\bar{\alpha})}_{\sigma, A, n}(x_{i} + a/\sqrt{2}, q_{y}),
\end{align}

\enni in the vicinity of $k_{y} \approx K_{Fy} + q_{y}$ with $q_{y}$ small. A similar expansion holds for $v_{A/B}$. Note that we are
considering bonding solutions where the envelope functions $u^{(\alpha)}_{A} = u^{(\alpha)}_{B}$, while keeping track of the phase difference between the two sublattices. Explicitly, for positive $K_{Fy}$, $\alpha = 1$ and $\bar{\alpha}=2'$, and $K^{(\alpha)}= K_{Fx}$ while $K^{(\bar{\alpha)}}= -K_{Fx}$. A similar 
expansion holds for valleys (2), and (1') with opposite signs for both $K_{Fx}$ and $K_{Fy}$. 
This form for the ansatz is clearly justified for zero strain, where the pairs of valleys $(1)$ and $(2')$ and $(1')$ and $(2)$ are decoupled. Any amount of small strain breaks translational symmetry along the $x$ direction and can in principle mix the valleys in pairs. The ansatz continues to hold, as indicated by numerical results which do not assume this decomposition (see main text).

Following the procedure in Sec.~\ref{Sec:IIb}, we formally pass to the continuum by expanding in terms of Wannier states. We also drop the $i$, $y$ and the sublattice index. Our ansatz becomes 

\enni \begin{align}
\gamma^{(\alpha)}_{s, n}(q) = & \sum_{\sigma} \int dx e^{-iK^{(\alpha)}_{Fx}x} \bigg\{ \left[ u^{(\alpha), *}_{s\sigma, n}(x, q) \Psi^{(I)}_{A\sigma}\left(x, q \right) + v^{(\alpha), *}_{s\sigma, n}(x, q) \Psi^{(II),\dag}_{A\sigma} \left(x, -q \right) \right] \notag \\
+ & e^{-iK^{(\alpha)}_{Fx}a/\sqrt{2}} \left[ u^{(\alpha), *}_{s\sigma,  n}(x+ a/\sqrt{2}, q) \Psi^{(I)}_{B, \sigma}\left(x, q \right) + v^{(\alpha), *}_{s\sigma, n}(x+a/\sqrt{2}, q) \Psi^{(II),\dag}_{B\sigma} \left(x, -q \right) \right] \bigg\},~\alpha \in \{1 , 2'\}.
\end{align}

\enni We introduced new field operators $\Psi^{(I/II),\dag}_{\sigma} \left(x, q \right)$ in the vicinity of $k_{y} \approx \pm K_{Fy} + q$. For valleys $1', 2$ we must interchange $I, II$ indices. This decomposition is equivalent to 

\enni \begin{align}
\gamma_{s, n}\left(K^{(\alpha)}_{Fy} + q \right) = \gamma^{(\alpha)}_{s, n}(q) + \gamma^{(\bar{\alpha})}_{s, n}(q), 
\end{align}

\enni With this ansatz, the 
BdG equations can be separated into valley sectors.

To illustrate the general solution, we consider valley (1) 
in particular. 
We expand the envelope functions in real space and the trigonometric functions in momentum space about the nodal points. We furthermore approximate $t(x_{i} + a/\sqrt{2})$ by $t(x_{i})$ since the effect of the strain is assumed to vary slowly on the scale of the lattice. We keep only terms which are first order in the small terms $\delta t(x), \partial_{x}, q$. To zeroth order we obtain:

\enni \begin{align}
\left[4t \cos \left( \frac{K_{Fx}a}{\sqrt{2}} \right) \cos \left( \frac{K_{Fy}a}{\sqrt{2}} \right)  \right]u^{(1)}_{s, n}(x, q) - \mu u^{(1)}_{s, n}(x, q) = 0. 
\end{align}

\enni To first order, we obtain

\begin{align}
& \left[ 4 \delta t(x) \cos \left( \frac{K_{Fx}a}{\sqrt{2}} \right) \cos \left( \frac{K_{Fy}a}{\sqrt{2}} \right) 
- 2 ta \sqrt{2} \cos \left( \frac{K_{Fx}a}{\sqrt{2}} \right) \sin \left( \frac{K_{Fy}a}{\sqrt{2}} \right)q + 2ta \sqrt{2} \cos \left( \frac{K_{Fy}a}{\sqrt{2}} \right) \sin \left( \frac{K_{Fx}a}{\sqrt{2}} \right) i \partial_{x} \right] u^{(1)}_{s, n}(x, q) \notag \\
& + \left[ 2 \Delta \sqrt{2} a \sin\left(K_{Fx}a\sqrt{2}\right) i \partial_{x} + 
2 \Delta a \sqrt{2} \sin\left(K_{Fy}a\sqrt{2}\right) q
\right]
 v^{(1)}_{s,n}(x, q) = E_{n}(q) u^{(1)}_{s, n}(x, q) 
\end{align}

We define 

\enni \begin{align}
v_{F} = & 2 ta \sqrt{2} \cos \left( \frac{K_{F}a}{\sqrt{2}} \right) \sin \left( \frac{K_{F}a}{\sqrt{2}} \right) \\
= & ta \sqrt{2} \sin\left(K_{F}a\sqrt{2}\right) \\
v_{\Delta} = & 2 \Delta a \sqrt{2} \sin\left(K_{F}a\sqrt{2}\right),\\
A_{0} = & 4t \cos^{2} \left( \frac{K_{F}a}{\sqrt{2}} \right), \\
A_{1}(x) = & 4\delta t(x) \cos^{2} \left( \frac{K_{F}a}{\sqrt{2}} \right)
\end{align}

\enni where $K_{F}= |K_{Fx}| = |K_{Fy}| \ge 0$. 

\enni Thd BdG eqs. simplify to

\enni \begin{align}
v_{F} \left[ i \partial_{x} -  q + \frac{A}{v_{F}}\right]u^{(1)}_{s, n}(x, q) 
+  v_{\Delta} \left[ i\partial_{x} + q \right] v^{(1)}_{s, n}(x, q) = & E_{n}(q) u^{(1)}_{s, n}(x, q) \notag \\
 v_{\Delta} \left[ i\partial_{x} +  q \right] u^{(1)}_{s, n}(x, q) - 
 v_{F} \left[ i \partial_{x} -  q + \frac{A}{v_{F}}\right] v^{(1)}_{s, n}(x, q) 
 = & E_{n}(q) v^{(1)}_{s, n}(x, q). 
 \label{Eq:Dgnl_1}
\end{align}

\enni We now focus on the zero-energy solutions.  We transform  $\tau_{z} \rightarrow \tau_{y}$ and obtain

\enni \begin{align}
\left\{ v_{\Delta} \left[ i\partial_{x} + q \right] - i v_{F} \left[ i \partial_{x} -  q + \frac{A_{1}}{v_{F}}\right] 
  \right\} \tilde{v}^{(1)}_{s, a, 0}(x, q) = 0 \label{Eq:Dgnl_hrmn_1}\\
\left\{ v_{\Delta} \left[ i\partial_{x} +  q \right] + 
 i v_{F} \left[ i \partial_{x} -  q + \frac{A_{1}}{v_{F}}\right] \right\} \tilde{u}^{(1)}_{s, a, 0}(x, q) 
 = & 0. \label{Eq:Dgnl_hrmn_2}
\end{align}

\enni These equations are very similar to those for the zeroth LL  obtained in Sec. II. By introducing the quantity 

\enni \begin{align}
v_{F} \pm i v_{\Delta} = v e^{\pm i \theta},
\end{align}

\enni we can write

\enni \begin{align}
\left[ \partial_{x} +i e^{-2i \theta} q-
\frac{i v_{F} e^{-i\theta} A_{1}(x)}{v} \right] \tilde{v}_{s, 0}(x, q) = & 0\\
\left[  \partial_{x} + i e^{2i\theta} q-\frac{i v_{F} e^{i\theta} A_{1}(x)}{v}\right] \tilde{u}_{s, 0}(x, q) = & 0. 
\end{align}

\enni The solutions are given by 

\enni \begin{align}
\tilde{v}_{s, 0}(x, q) = & C_{v} \exp\left\{ \frac{\left[\sin(2\theta) q-
\sin(\theta) \frac{ v_{F} A_{1}(x)}{v} \right]^{2}}{2\frac{v_{F}}{v_{\Delta}}\sin(\theta)\partial_{x}A_{1}(x)} \right\}
 \exp\left\{ i\frac{\left[\cos(2\theta) q-
\cos(\theta) \frac{ v_{F} A_{1}(x)}{v} \right]^{2}}{2\frac{v_{F}}{v_{\Delta}}\cos(\theta)\partial_{x}A_{1}(x)} \right\} \\
\tilde{u}_{s, 0}(x, q) = & C_{u} \exp\left\{ - \frac{\left[\sin(2\theta) q-
\sin(\theta) \frac{ v_{F} A_{1}(x)}{v} \right]^{2}}{2\frac{v_{F}}{v_{\Delta}}\sin(\theta)\partial_{x}A_{1}(x)} \right\}
 \exp\left\{ i\frac{\left[\cos(2\theta) q-
\cos(\theta) \frac{ v_{F} A_{1}(x)}{v} \right]^{2}}{2\frac{v_{F}}{v_{\Delta}}\cos(\theta)\partial_{x}A_{1}(x)} \right\}
\end{align}

\enni  We recall that $A(x) \sim x$ by construction. Therefore, only \emph{one} solution 
is normalizable since it contains a decaying Gaussian form. This is typical of the zeroth LL wavefunctions for Dirac systems in a Landau gauge.

Without loss of generality, we consider the case where $\tilde{u}$ is the allowed non-trivial solution:

\enni \begin{align}
\begin{pmatrix}
u^{(1)}_{s, 0} \\
v^{(1)}_{s, 0}
\end{pmatrix}
= & \frac{1}{\sqrt{2}} \tilde{u}(x, q)
\begin{pmatrix}
1 \\
- i
\end{pmatrix}.
\label{Eq:Dgnl_sltn_1}
\end{align}

The equations for valley (2') can be 
obtained from Eqs.~\ref{Eq:Dgnl_1} by changing the sign of $\partial_{x}$. The analogues of Eqs.~\ref{Eq:Dgnl_hrmn_1}-\ref{Eq:Dgnl_hrmn_2} involve complex conjugation and interchanging $\tilde{u}$ and $\tilde{v}$: 

\enni \begin{align}
\begin{pmatrix}
u^{(2')}_{s, 0}(x, q) \\
v^{(2')}_{s, 0}(x, q) 
\end{pmatrix}
= e^{i \phi^{(2')}_{s}}
\begin{pmatrix}
-v^{(1), *}_{s, 0}(x, q) \\
u^{(1), *}_{s, 0}(x, q)
\end{pmatrix}.
\end{align}

\enni We introduced a global phase term, which must be chosen 
to ensure consistency with other symmetries. We can readily check that these are consistent with the mirror symmetry given by Eq.~\ref{Eq:Mrr_smmt}. Explicitly, we have 

\enni \begin{align}
\begin{pmatrix}
u^{(2')}_{s, 0}(x, q) \\
v^{(2')}_{s, 0}(x, q)
\end{pmatrix}
= \frac{\text{sgn}(s)}{\sqrt{2}}\tilde{u}^{*}(x, q)
\begin{pmatrix}
-i  \\
1
\end{pmatrix},
\label{Eq:Dgnl_sltn_2}
\end{align} 

\enni where we fixed the global phase.

The BdG eqs. at valleys (1') can be obtained from Eqs.~\ref{Eq:Dgnl_1} by changing the signs of both $\partial_{x}$ and $q$. The solutions are given by complex-conjugating, interchanging $\tilde{u}$ and $\tilde{v}$ and \emph{changing the sign of $q$}. These steps are nothing 
but the p-h transformation in Eq.~\ref{Eq:Sngl_ph_trns}. The explicit solutions read

\enni \begin{align}
\begin{pmatrix}
u^{(1')}_{s, 0}(x, q) \\
v^{(1')}_{s, 0}(x, q)
\end{pmatrix}
= \frac{i}{\sqrt{2}}\tilde{u}^{*}(x, -q)
\begin{pmatrix}
-i  \\
1
\end{pmatrix},
\end{align} 

\enni and 

\enni \begin{align}
\begin{pmatrix}
u^{(2)}_{s, 0}(x, q) \\
v^{(2)}_{s, 0}(x, q)
\end{pmatrix}
= \frac{-i \text{sgn}(s)}{\sqrt{2}}\tilde{u}(x, -q)
\begin{pmatrix}
-1  \\
i
\end{pmatrix}.
\end{align} 

\enni Note that we allowed for additional global phases $i$ to ensure time-reversal symmetric solutions at opposite valleys. This does not violate the p-h correspondence since solutions are determined modulo a global phase.

To sum up the discussion thus far, we obtain 
two independent solutions for $k_{y}\approx K_{Fy}+ q_{y}$, where $K_{Fy}$ is positive:

\enni \begin{align}
\gamma^{(1)}_{s, 0}(q) = & \sum_{\sigma} \int dx e^{-iK_{Fx}x} \bigg\{ \tilde{u}^{*}(x,q) \left[ \delta_{s\sigma} \Psi^{(I)}_{A\sigma}\left(x, q \right) + i (i \sigma_{y})_{s\sigma} \Psi^{(II),\dag}_{A\sigma} \left(x, -q \right) \right] \notag \\
+ & e^{-iK_{Fx}a/\sqrt{2}} \tilde{u}^{*}(x+a/\sqrt{2}, q) \left[ \delta_{s\sigma} \Psi^{(I)}_{B, \sigma}\left(x, q \right) + i (i \sigma_{y})_{s\sigma} \Psi^{(II),\dag}_{B\sigma} \left(x, -q \right) \right] \bigg\} \\
& \notag \\
\gamma^{(2')}_{s, n}(q) = & \text{sgn}(s) \sum_{\sigma} \int dx e^{iK_{Fx}x} \bigg\{ \tilde{u}(x,q) \left[ i \delta_{s\sigma} \Psi^{(I)}_{A\sigma}\left(x, q \right) + (i \sigma_{y})_{s\sigma} \Psi^{(II),\dag}_{A\sigma} \left(x, -q \right) \right] \notag \\
+ & e^{iK_{Fx}a/\sqrt{2}} \tilde{u}(x+a/\sqrt{2}, q) \left[ i \delta_{s\sigma} \Psi^{(I)}_{B, \sigma}\left(x, q \right) + (i \sigma_{y})_{s\sigma} \Psi^{(II),\dag}_{B\sigma} \left(x, -q \right) \right] \bigg\} \\
& \notag \\
\gamma^{(1')}_{s, 0}(q) = & \sum_{\sigma} \int dx e^{iK_{Fx}x} \bigg\{ \tilde{u}(x,-q) \left[  \delta_{s\sigma} \Psi^{(II)}_{A\sigma}\left(x, q \right) -i(i\sigma_{y})_{s\sigma} \Psi^{(I),\dag}_{A\sigma} \left(x, -q \right) \right] \notag \\
+ & e^{iK_{Fx}a/\sqrt{2}} \tilde{u}(x+a/\sqrt{2}, -q) \left[  \delta_{s\sigma} \Psi^{(II)}_{B, \sigma}\left(x, q \right) -i(i \sigma_{y})_{s\sigma} \Psi^{(I),\dag}_{B\sigma} \left(x, -q \right) \right] \bigg\} \\
& \notag \\
\gamma^{(2)}_{s, 0}(q) = & \text{sgn}(s)\sum_{\sigma} \int dx e^{-iK_{Fx}x} \bigg\{ \tilde{u}^{*}(x,-q)  \left[ -i \delta_{s\sigma} \Psi^{(II)}_{A\sigma}\left(x, q \right) + (i\sigma_{y})_{s\sigma} \Psi^{(I),\dag}_{A\sigma} \left(x, -q \right) \right] \notag \\
+ & e^{-iK_{Fx}a/\sqrt{2}} \tilde{u}^{*}(x+a/\sqrt{2}, -q) \left[  -i \delta_{s\sigma} \Psi^{(II)}_{B, \sigma}\left(x, q \right)  + (i\sigma_{y})_{s\sigma} \Psi^{(I),\dag}_{B\sigma} \left(x, -q \right) \right] \bigg\}
\end{align}

\enni These satisfy

\enni \begin{align}
\gamma^{(1')}_{s', 0}(q) = &  \sum_{s} (\sigma_{y})_{s's} \gamma^{(1), \dag}_{s, 0}(-q) \label{Eq:Dgnl_Mjrn_cndt_1} \\
\gamma^{(2)}_{s', 0}(q) = &  \sum_{s} - (\sigma_{y})_{s's} \gamma^{(2'), \dag}_{s, 0}(-q). \label{Eq:Dgnl_Mjrn_cndt_2}
\end{align}

\enni These imply that at each valley there are two independent solutions. However, solutions at opposite valleys are not independent, in analogy with the results of Sec.~\ref{Sec:IIc}.

\subsection{Comparison to numerical solution of the lattice model}
\label{Sec:IIIc}

In order to compare with the results of the numerical calculation presented in Fig.~4 of the main text, we consider the linear combinations of the analytical solutions found in the previous section

\enni \begin{align}
\gamma^{(I)}_{s, 1, 0}(q) = & \gamma^{(1)}_{s, 0}(q) - i\text{sgn}(s) \gamma^{(2')}_{s, 0}(q) \\
\gamma^{(I), \dag}_{s, 2, 0}(q) = & -i \gamma^{(1)}_{s, 0}(q) + \text{sgn}(s)  \gamma^{(2')}_{s, 0}(q), \\
\end{align}

\enni which imply 


\enni \begin{align}
\gamma^{(I)}_{1, s, 0}(q) = & \int dx \left[ u^{(I)}_{1, A, s \sigma}\Psi^{(I)}_{A\sigma}(x, q)  + v^{(I)}_{1, A, s \sigma} \Psi^{(II)\dag}_{A\sigma}(x, -q) \right] + \left[ u^{(I)}_{1, B, s \sigma}\Psi^{(I)}_{B\sigma}(x, q)  + v^{(I)}_{1, B, s \sigma} \Psi^{(II)\dag}_{B\sigma}(x, -q) \right]\\
\gamma^{(I)}_{2, s, 0}(q) = & \int dx \left[ u^{(I)}_{2, A, s \sigma}\Psi^{(I)}_{A\sigma}(x, q)  + v^{(I)}_{2, A, s \sigma} \Psi^{(II)\dag}_{A\sigma}(x, -q) \right] + \left[ u^{(I)}_{2, B, s \sigma}\Psi^{(I)}_{B\sigma}(x, q)  + v^{(I)}_{2, B, s \sigma} \Psi^{(II)\dag}_{B\sigma}(x, -q) \right],
\end{align}

\enni where

\enni \begin{align}
u^{(I)}_{1, A, s \sigma} = & 2 \text{Re}\left[ e^{iKx} \tilde{u} (x, q)\right] \delta_{s\sigma}  \\
v^{(I)}_{1, A, s \sigma} = &  2 \text{Im}\left[ e^{iKx} \tilde{u} (x, q)\right] (i\sigma_{y})_{s\sigma} \\
u^{(I)}_{1, B, s \sigma} = & 2 \text{Re}\left[ e^{iK(x+a/\sqrt{2}} \tilde{u} (x, q)\right] \delta_{s\sigma}  \\
v^{(I)}_{1, B, s \sigma} = &  2 \text{Im}\left[ e^{iK(x+ a/\sqrt{2}} \tilde{u} (x, q)\right] (i\sigma_{y})_{s\sigma} \\
& \notag \\
u^{(I)}_{2, A, s \sigma} = & - 2 \text{Im}\left[ e^{iKx} \tilde{u} (x, q)\right] \delta_{s\sigma}  \\
v^{(I)}_{2, A, s \sigma} = &  2 \text{Re}\left[ e^{iKx} \tilde{u} (x, q)\right] (i\sigma_{y})_{s\sigma} \\
u^{(I)}_{2, B, s \sigma} = & - 2 \text{Im}\left[ e^{iK(x+a/\sqrt{2}} \tilde{u} (x, q)\right] \delta_{s\sigma}  \\
v^{(I)}_{2, B, s \sigma} = &  2 \text{Re}\left[ e^{iK(x+ a/\sqrt{2}} \tilde{u} (x, q)\right] (i\sigma_{y})_{s\sigma}.
\end{align}

\enni Similar linear combinations can be taken wrt valley (II) operators. These can be obtained via a time-reversal operation

\enni \begin{align}
\sum_{s'} (-i \sigma_{s})ss' \gamma^{(II)}_{1, s', 0}(-q) =  \theta\gamma^{(I)}_{1, s, 0}(q),
\end{align}

\enni which gives

\enni \begin{align}
\begin{pmatrix}
u^{(II)}_{1/2, A/B, s}(q) \\
v^{(II)}_{1/2, A/B, s}(q)
\end{pmatrix}
= (-i \sigma_{y})_{ss'} 
\begin{pmatrix}
u^{(I)}_{1/2, A/B, s'}(-q) \\
v^{(I)}_{1/2, A/B, s'}(-q),
\end{pmatrix}
\end{align}

\enni where $u^{(\alpha)}_{s\sigma} = \delta_{s\sigma} u^{(\alpha)}_{s}$ and $v^{(\alpha)}_{s\sigma} = (i\sigma_{y})_{s\sigma} v^{(\alpha)}_{s}$. Hence, the BdG coefficients can be made real.

One can see that these obey the constraints imposed by the mirror symmetry discussed in the main text. Furthermore, these analytical solutions are consistent with the numerical results of Fig.~4 of the main text.

%

\subsection{Projection onto zeroth LLs}
\label{Sec:IIId}

The inverse BdG eqs. can be obtained via generalizing Eqs.~\ref{Eq:Sngl_invr_BdG} to the case of two suballtices:

\enni \begin{align} 
\Psi^{(I)}_{A, \sigma}\left(x, q \right)= & \sum_{n, s} e^{i K^{(\alpha)}_{Fx}x} u^{(\alpha)}_{ A, s\sigma, n}(x, q) \gamma^{(\alpha)}_{s, n}(q) + 
e^{-iK^{(\bar{\alpha})}_{Fx}x}v^{(\bar{\alpha}),*}_{A, s \sigma, n}(x, -q) \gamma^{(\bar{\alpha}),\dagger}_{s, n}(-q), ~\alpha \in \{1, 2'\}  \\
\Psi^{(I)}_{B, \sigma}\left(x, q \right)= & \sum_{n, s} e^{i K^{(\alpha)}_{Fx}x} u^{(\alpha)}_{ B, s\sigma, n}(x, q) \gamma^{(\alpha)}_{s, n}(q) + 
e^{-i K^{(\alpha)}_{Fx}x} v^{(\bar{\alpha}),*}_{B, s \sigma, n}(x, -q) \gamma^{(\bar{\alpha}),\dagger}_{s, n}(-q),~\alpha \in \{1, 2'\}  \\
& \\
\Psi^{(II)}_{A, \sigma}\left(x, q \right)= & \sum_{n, s} e^{i K^{(\alpha)}_{Fx}x} u^{(\alpha)}_{ A, s\sigma, n}(x, q) \gamma^{(\alpha)}_{s, n}(q) + 
e^{-iK^{(\bar{\alpha})x}_{Fx}}v^{(\bar{\alpha}),*}_{A, s \sigma, n}(x, -q) \gamma^{(\bar{\alpha}),\dagger}_{s, n}(-q),~\alpha \in \{1', 2\} \\
\Psi^{(II)}_{B, \sigma}\left(x, q \right)= & \sum_{n, s} e^{i K^{(\alpha)}_{Fx}x} u^{(\alpha)}_{ B, s\sigma, n}(x, q) \gamma^{(\alpha)}_{s, n}(q) + 
e^{-i K^{(\alpha)}_{Fx}x} v^{(\bar{\alpha}),*}_{B, s \sigma, n}(x, -q) \gamma^{(\bar{\alpha}),\dagger}_{s, n}(-q),~\alpha \in \{1', 2\}.
\end{align} 

As discussed in Sec.~\ref{Sec:IId}, we truncate the Hilbert space by considering only $(1)$ and $(2')$ solutions. This ensures that the  BdG 
transformation on the zeroth LL sector is well-defined are well-defined. The projected operators are given by 

\enni \begin{align}
\Psi^{(I)}_{P, A \up}(x, q) = & e^{iK_{F}x}u^{(1)}_{A, \up, 0}(x, q) \gamma^{(1)}_{\up, 0}(q) 
+ e^{-iK_{F}x}u^{(2')}_{A, \up, 0}(x, q) \gamma^{(2')}_{\up, 0}(q) \\
\Psi^{(I)}_{P, A \dn}(x, q) = & e^{iK_{F}x}u^{(1)}_{A, \dn, 0}(x, q) \gamma^{(1)}_{\dn, 0}(q) 
+ e^{-iK_{F}x}u^{(2')}_{A, \dn, 0}(x, q) \gamma^{(2')}_{\dn, 0}(q) \\
\Psi^{(II)}_{P, A \up}(x, q) = & - e^{-iK_{F}x}v^{(1), *}_{A, \up, 0}(x, -q) \gamma^{(1), \dag}_{\dn, 0}(-q) 
- e^{iK_{F}x}v^{(2'), *}_{A, \up, 0}(x, -q) \gamma^{(2'), \dag}_{\dn, 0}(-q) \\
\Psi^{(II)}_{P, A \dn}(x, q) = &  e^{-iK_{F}x}v^{(1), *}_{A, \dn, 0}(x, -q) \gamma^{(1), \dag}_{\up, 0}(-q) 
+ e^{iK_{F}x}v^{(2'), *}_{A, \dn, 0}(x, -q) \gamma^{(2')}_{\up, 0}(-q),
\end{align}

\enni with similar expressions for $B$. Using Eqs.~\ref{Eq:Dgnl_sltn_1},~\ref{Eq:Dgnl_sltn_2}, it is straightforward to verify that 

\enni \begin{align}
\Psi^{(II)}_{P, A \sigma}(x, q) = & (\sigma_{y})_{\sigma \sigma'} \Psi^{(I), \dag}_{P, A \sigma}(x, -q).
\label{Eq:Dgnl_Mjrn}
\end{align}

\enni Similar relations hold for sublattice $B$. These expressions are the analogues of Eq.~\ref{Eq:Sngl_prjc_Mjrn}. 

Following the procedure set out in Sec.~\ref{Sec:IId}, we consider the real space fields on each sublattice:

\enni \begin{align}
\Psi^{(I)}_{A/B,\sigma}(x, y) = &  \int \left( \frac{dq}{2\pi} \right) e^{i q y}\Psi^{(I)}_{A/B \sigma}(x, q) \\
\Psi^{(II)}_{A/B,\sigma}(x, y) = &  \int \left( \frac{dq}{2\pi} \right) e^{i q y}\Psi^{(II)}_{A/B \sigma}(x, q).
\end{align}

Finally, the real-space fields are determined by 

\enni \begin{align}
\Psi_{A/B \sigma}(x, y) = e^{iK_{Fy}a} \Psi^{(I)}_{A/B,\sigma}(x, y) + e^{-iK_{Fy}a} \Psi^{(II)}_{A/B,\sigma}(x, y).
\end{align}

\enni Note that the overall phases for momenta along the $y$ direction.

\subsection{Ferromagnetism at mean-field level}
\label{Sec:IIIe}

As for the case of strain along a nodal axis, we allow small but finite residual Hubbard interactions. In contrast to the previous case, there are now two independent states per spin at each valley, related to each other by Eqs.~\ref{Eq:Dgnl_Mjrn_cndt_1},~\ref{Eq:Dgnl_Mjrn_cndt_2}. As discussed in Sec.~\ref{Sec:IIIa} the real-space system is described by a two-site unit cell. The two independent zero-energy LLs per spin at each valley can be used to determine the projections of the two fields $\Psi_{P, A \sigma}$ and $\Psi_{P, B, \sigma}$ defined on each sublattice, as shown in the previous section. 

Consider a continuum version of the Hubbard interaction on each 
sublattice:

\enni \begin{align}
H_{U, P} = & U' \int d^{2}r \left[ \Psi^{\dagger}_{A \up}(\bm{r})\Psi_{A \up}(\bm{r})\Psi^{\dagger}_{A \dn}(\bm{r})\Psi_{A \dn}(\bm{r}) +
\Psi^{\dagger}_{B \up}(\bm{r})\Psi_{B \up}(\bm{r})\Psi^{\dagger}_{B \dn}(\bm{r})\Psi_{B \dn}(\bm{r}) \right].
\end{align}

\enni Using Eqs.~\ref{Eq:Dgnl_Mjrn}, each sublattice sector maps onto the single-valley case studied in Sec.~\ref{Sec:IIe}. Hence, each sublattice is expected to order ferromagnetically at mean-field level.

The large degeneracy due to possible relative orientations of the resulting sublattice moments can be lifted by arbitrarily small inter-sublattice interactions. In view of expected short range, repulsive nature of the residual interactions in a SC, and in the absence of negligible dispersion, we conclude that the most likely instability is still ferromagnetic. 

\section{Helical Majoranas for 3D equal-spin triplet SCs}
\label{Sec:IV}

In this Section we consider the emergence of helical Majorana modes in 3D equal-spin triplet superconductors under uniaxial strain. Sec.~\ref{Sec:IVa} introduces a family of effective 1D lattice models. In Sec.~\ref{Sec:IVb}, we obtain helical Majorana states in the continuum limit. In Sec.~\ref{Sec:IVc} we show that the Fermi fields projected onto the zeroth LLs are emergent real-space Majorana fermions.

\subsection{Lattice Hamiltonian}
\label{Sec:IVa}

We consider a 3D lattice model with NN hopping along either $x, y, z$ directions and strain along the $x$ directions. The pairing is chosen to belong to the $\Gamma^{-}_{1}$ representation of the tetragonal $D_{4h}$ point group~\cite{Sigrist}. Without strain, this can be written as 

\enni \begin{align}
H= \sum_{\bm{k}}
\begin{pmatrix}
c^{\dagger}_{\bm{k} \uparrow} \\
c^{\dagger}_{\bm{k} \downarrow} \\
c_{- \bm{k} \downarrow} \\
- c_{- \bm{k} \uparrow} 
\end{pmatrix}^{T} 
\begin{pmatrix}
h_{\bm{k}} & 0 & 0 & - \Delta_{\bm{k}} \\
0 & h_{\bm{k}} & \Delta'_{\bm{k}} & 0  \\
0 & \Delta^{', *}_{\bm{k}} &  - h_{\bm{k}} & 0 \\
- \Delta^{*}_{\bm{k}} &  0 & 0 & -h_{\bm{k}}
\end{pmatrix}
\begin{pmatrix}
c_{\bm{k} \uparrow} \\
c_{\bm{k} \downarrow} \\
c^{\dagger}_{- \bm{k} \downarrow} \\
- c^{\dagger}_{- \bm{k} \uparrow}
\end{pmatrix},
\label{Eq:3D_prst_Hmlt}
\end{align}

\enni where 

\enni \begin{align}
h_{\bm{k}} = & 2t \left( \cos(k_{x}a) + \cos(k_{y}a) + \cos(k_{z}a) \right) - \mu \\
\Delta_{\bm{k}} = & \Delta \left( - \sin(k_{x}a) + i \sin(k_{y}a) \right) \\
\Delta^{'}_{\bm{k}} = & \Delta \left( \sin(k_{x}a) + i \sin(k_{y}a) \right).
\end{align}

\enni The spectrum is given by 

\enni \begin{align}
E_{\bm{k}} = & \pm \sqrt{h^{2}_{\bm{k}}+ \Delta^{2} \left(\sin(k_{x}a)^{2} + \sin(k_{y}a)^{2} \right) },
\end{align}

\enni with Dirac point nodes at $\bm{K}_{1/1'}= (0, 0, \pm K_{F})$. 

The lattice Hamiltonian can be chosen as

\enni \begin{align}
H= H_{TB} + H_{\text{Pair}}
\end{align}

\enni where 

\enni \begin{align}
H_{TB} = \sum_{\bm{R}_{i}} \sum_{\bm{\delta}_{j}} \sum_{\sigma} t(x_{i}) c^{\dagger}_{\sigma}(\bm{R}_{i}) c_{\sigma}(\bm{R}_{i}+ \bm{\delta}_{j}) + \text{H.c.} - \mu c^{\dagger}_{\sigma}(\bm{R}_{i}) c_{\sigma}(\bm{R}_{i}).
\end{align}

\enni where $\bm{\delta}_{j=x, y, z}$ equals $(a, 0, 0), (0, a, 0), (0, 0, a)$, respectively. The pairing part is given by

\enni \begin{align}
H_{\text{Pair}} = & \frac{\Delta}{2} \sum_{x_{i}, y_{j}, z_{l}} \sum_{\sigma} \bigg\{i \text{sgn}(\sigma) \left[  c^{\dagger}_{\sigma}(x_{i}, y_{j}, z_{l}) c^{\dagger}_{\sigma}(x_{i+1}, y_{j}, z_{l}) - 
c^{\dagger}_{\sigma}(x_{i}, y_{j}, z_{l}) c^{\dagger}_{\sigma}(x_{i-1}, y_{j}, z_{l}) \right] \notag \\
+ & \left[ c^{\dagger}_{\sigma}(x_{i}, y_{j}, z_{l}) c^{\dagger}_{\sigma}(x_{i}, y_{j+1}, z_{l}) - 
c^{\dagger}_{\sigma}(x_{i}, y_{j}, z_{l}) c^{\dagger}_{\sigma}(x_{i}, y_{j-1}, z_{l}) \right] \bigg\} + \text{H.c.}
\end{align}

\enni Note the difference in sign due to the convention in Eq.~\ref{Eq:3D_prst_Hmlt}. 

We apply Fourier transforms along the $y$ and $z$ directions. As in the pristine case, the Hamiltonian, and consequently the BdG equations separate into two independent sectors:

\enni \begin{align}
t(x_{i}) \left[u_{s, n}(x_{i+1}, \bm{k})  +  u_{s, n}(x_{i-1}, \bm{k}) \right] + \left\{ 2t(x_{i}) \left[ \cos(k_{y}a)  +
\cos(k_{z}a) \right] - \mu \right\}u_{s , n}(x_{i}, \bm{k}) \notag \\
- \frac{i \Delta}{2}  \left[ \left(v_{s, n}(x_{i+1}, \bm{k}) - v_{s, n}(x_{i-1}, \bm{k}) \right) + 2 \text{sgn}(\sigma) \sin(k_{y}a) v_{s, n}(x_{i}, \bm{k}) \right] = E_{n}(\bm{k}) u_{s, n}(x_{i}, \bm{k}) \label{Eq:Trpl_BdG_1} \\
 \frac{i \Delta}{2}  \left[ \left(u_{s, n}(x_{i-1}, \bm{k}) - u_{s, n}(x_{i+1}, \bm{k}) \right) + 2 \text{sgn}(s) \sin(k_{y}a) u_{s, n}(x_{i}, \bm{k}) \right] \notag \\
 - t(x_{i}) \left[v_{s, n}(x_{i+1}, \bm{k})  +  v_{s, n}(x_{i-1}, \bm{k}) \right] - \left\{ 2t(x_{i}) \left[ \cos(k_{y}a)  +
\cos(k_{z}a) \right] -\mu \right\} v_{s, n}(x_{i}, \bm{k}) = E_{n}(\bm{k}) v_{s, n}(x_{i}, \bm{k}). \label{Eq:Trpl_BdG_2}
\end{align}

\enni In contrast to the singlet cases of Sec.~\ref{Sec:II} and~\ref{Sec:III}, the BdG eqs. for each $H_{s}$ retain a dependence on spin index $s$.

\subsection{Landau levels in the continuum limit}
\label{Sec:IVb}

We proceed as in Sec.~\ref{Sec:IIb} and consider ansatze 

\enni \begin{align}
u_{s, n}(x_{i}, k_{y}, k_{z} ) \approx e^{iK^{(\alpha)_{F}}z_{l}} 
u^{(\alpha)}_{s, n}(x_{i}, k_{y}, k_{z} ) + e^{iK^{(\bar{\alpha})_{F}}z_{l}} 
u^{(\bar{\alpha})}_{s, n}(x_{i}, k_{y}, k_{z} ),
\label{Eq:Trpl_anst}
\end{align}

\enni for the wavefunctions in the vicinity of the two valleys located at $\bm{K} = (0, 0 \pm K_{F})$ and labeled by $(1)$ and $(1')$. A similar expansion is expected to hold for the $v$'s. With this ansatz, the BdG eqs. for each $H_{s}$ further separate into two valley sectors. We proceed by taking the continuum limit. Since the envelope functions in Eq.~\ref{Eq:Trpl_anst} are assumed to vary slowly on the scale of the lattice, we can expand 

\enni \begin{align}
u^{(\alpha)}_{s, n}\left(x+a, q_{y}, K^{(\alpha)}_{F} + q_{z}\right) + u^{(\alpha)}_{s, n}\left(x-a, q_{y}, K^{(\alpha)}_{F} + q_{z}\right) \approx \notag \\
 \approx u^{(\alpha)}_{s, n}\left(x, q_{y}, q_{z}\right) + a \partial_{x} u^{(\alpha)}_{s, n}\left(x, q_{y}, q_{z}\right) 
+ u^{(\alpha)}_{s, n}\left(x, q_{y}, q_{z}\right) - a \partial_{x} u^{(\alpha)}_{s, n}\left(x, q_{y}, q_{z}\right) \notag \\
= 2 u^{(\alpha)}_{s, n}\left(x, q_{y}, q_{z}\right).
\end{align}

\enni Similarly, we have 

\begin{align}
v^{(\alpha)}_{s, n}(x+a, q_{y}, q_{z}) - v^{(\alpha)}_{s, n}(x-a, q_{y}, q_{z}) \approx 
2a \partial_{x} v^{(\alpha)}_{s, n}(x, q_{y}, q_{z}).
\end{align}

\enni We can likewise expand to first order in $q_{y}, q_{z}$

\enni \begin{align}
\cos\left((K_{Fy}+ q_{y}) a\right) = & 1 \\
\cos\left((K_{Fz}+ q_{z})a\right) = & \cos\left(K_{F}a\right) - a \text{sgn}\left(K_{Fz}\right) \sin\left(K_{F}\right) q_{z} \\
\sin\left(K_{Fy}+ q_{y}\right) = & a q_{y},
\end{align}

\enni since $K_{Fy}=0$.

\enni Allowing for $t(x)=t + \delta t(x)$, we write Eq.~\ref{Eq:Trpl_BdG_1} by retaining terms to first order in either $q_{y}, \partial_{x}, \delta t(x)$:

\enni \begin{align}
\left\{ \delta t(x)\left[4+ 2\cos\left(K_{F}a \right)  \right]  - 2t a \text{sgn}\left(K_{Fz}\right) \sin\left(K_{F}\right) q_{z} \right\}u^{(\alpha)}_{s, n}\left(x, q_{y}, q_{z}\right) 
+ \left[ - i a \Delta \partial_{x} - i a \Delta \text{sgn}(s) q_{y} \right]v^{(\alpha)}_{s, n}\left(x, q_{y}, q_{z}\right) \notag \\
 = E_{n}(q_{y}, q_{z}) u^{(\alpha)}_{s, n}\left(x, q_{y}, q_{z}\right), 
\end{align}

\enni with a similar expression for Eq.~\ref{Eq:Trpl_BdG_2}.
We define 

\enni \begin{align}
A(x) = & \frac{\delta t(x)\left[4+ 2\cos\left(K_{F}a \right)  \right]}{e} \\
v_{F} = & 2t a \sin\left(K_{F}\right) \\
v_{\Delta} = & a \Delta.
\end{align}

\enni The BdG eqs. can be written as

\enni \begin{align}
v_{F} \left[-\text{sgn}\left(K_{Fz}\right) q_{z} + \frac{eA}{v_{F}}\right]u^{(\alpha)}_{s, n}\left(x, q_{y}, q_{z}\right) - i v_{\Delta} \left[ \partial_{x} + \text{sgn}(s) q_{y}\right]v^{(\alpha)}_{s, n}\left(x, q_{y}, q_{z}\right) = & E_{n}(q_{y}, q_{z}) u^{(\alpha)}_{s, n}\left(x, q_{y}, q_{z}\right) \\
- i v_{\Delta} \left[ \partial_{x} - \text{sgn}(s) q_{y}\right]u^{(\alpha)}_{s, n}\left(x, q_{y}, q_{z}\right) - v_{F} \left[-\text{sgn}\left(K_{Fz}\right) q_{z} + \frac{eA}{v_{F}}\right]v^{(\alpha)}_{s, n}\left(x, q_{y}, q_{z}\right) 
= & E_{n}(q_{y}, q_{z}) v^{(\alpha)}_{s, n}\left(x, q_{y}, q_{z}\right), 
\end{align}

\enni or 

\enni \begin{align}
H^{(\alpha)}_{s} \Psi^{(\alpha)}_{\sigma, n} = & E_{n}(q_{y}, q_{z}) \Psi^{(\alpha)}_{s, n}, \\ 
H^{(\alpha)}_{s} = & v_{F} \left[-\text{sgn}\left(K^{(\alpha)}_{Fz}\right) q_{z} + \frac{eA}{v_{F}}\right] \tau_{z} +(-i \partial_{x}) \tau_{x} + v_{\Delta} \text{sgn}(s) q_{y} \tau_{y} 
\end{align}

%
%
%
%

Let us consider the solution at valley $(1)$. We assume that $\delta t(x)$ is chosen to generate a positive pseudo-magnetic field. As in Sec.~\ref{Sec:IIc}, we apply the transformation in Eq.~\ref{Eq:Trns_sngl} s.t. $\sigma_{z} \rightarrow \sigma_{y}$. The BdG Hamiltonian becomes

\enni \begin{align}
\tilde{H}^{(1)}_{s} = & v_{F} \left[- q_{z} + \frac{eA}{v_{F}}\right] \tau_{y} +v_{\Delta}(-i \partial_{x}) \tau_{x} - v_{\Delta} \text{sgn}(s) q_{y} \tau_{z} \notag \\
= & - v_{\Delta} \text{sgn}(s) q_{y} \tau_{z} + v_{\Delta}\left\{ (-i \partial_{x}) \tau_{x} + \left[-\lambda q_{z} + \frac{eA}{v_{\Delta}}\right] \tau_{y} \right\}.
\end{align}

\enni Following Ref.~\onlinecite{Balatsky}, we can choose the positive-energy solutions as

\enni \begin{align}
E_{n}(q_{y}, q_{z}) = \sqrt{\left( \omega_{c} \sqrt{n} \right)^{2} + 
\left(v_{\Delta} q_{z} \right)^{2}},~n = 0, 1, \hdots \\
\tilde{\Psi}^{(1)}_{s, n} = \text{sgn}(s)
\begin{pmatrix}
-i \alpha_{n} \psi_{n-1}\left[x, \lambda q_{z} \right] \\
\beta_{n} \psi_{n}\left[x, \lambda q_{z} \right]
\end{pmatrix},
\end{align}

\enni where $\psi$'s are the normalized harmonic oscillator wavefunctions and the coefficients are given by 

\enni \begin{align}
\alpha_{n} = & \sqrt{\frac{E_{n} - v_{\Delta} \text{sgn}(s) q_{y} }{2E_{n}}} \\
\beta_{n} = & \sqrt{\frac{E_{n} + v_{\Delta} \text{sgn}(s) q_{y} }{2E_{n}}}.
\end{align}

\enni The overall phase $\text{sgn}(s)$ is added to ensure that the solutions 
obey time-reversal symmetry. In the un-rotated basis the solutions are

\enni \begin{align}
\Psi^{(1)}_{s, n} = \text{sgn}(s)
\frac{1}{\sqrt{2}}
\begin{pmatrix}
-i \alpha_{n} \psi_{n-1}\left[x, \lambda q_{z} \right] - i \beta_{n} \psi_{n}\left[x, \lambda q_{z} \right] \\
- \alpha_{n} \psi_{n-1}\left[x, \lambda q_{z} \right]
+ \beta_{n} \psi_{n}\left[x, \lambda q_{z} \right]
\end{pmatrix}.
\end{align}

\enni For the zeroth LL, we have 

\enni \begin{align}
E_{0}(q_{y}, q_{z}) = v_{\Delta} \text{sgn}(q_{y}) q_{y},
\end{align}

\enni and we require $\beta_{0} =1$.  
The solutions are given by

\enni \begin{align}
\begin{pmatrix}
u^{(1)}_{s, 0}(x, q_{y}, q_{z}) \\
v^{(1)}_{s, 0} (x, q_{y}, q_{z} )
\end{pmatrix} 
= \frac{\text{sgn}(s)\psi_{0}(x, q )}{\sqrt{2}}
\begin{pmatrix}
-i \\
1
\end{pmatrix},~\text{sgn(s)}q_{y} \ge 0. 
\end{align}

\enni Therefore, the chiral spin-up zeroth LL mode is right-moving while it's spin-down counterpart is left-moving. 

We apply a similar procedure for valley $(1')$. Namely, 
we apply the inverse transformation which maps $\sigma_{z} \rightarrow -\sigma_{y}$ to get 

\enni \begin{align}
\tilde{H}^{(1')}_{s} = & -v_{F} \left[ q_{z} + \frac{eA}{v_{F}}\right] \tau_{y} +v_{\Delta}(-i \partial_{x}) \tau_{x} + v_{\Delta} \text{sgn}(s) q_{y} \tau_{z} \notag \\
= &  v_{\Delta} \text{sgn}(s) q_{y} \tau_{z} + v_{\Delta}\left\{ (-i \partial_{x}) \tau_{x} + \left[- \lambda q_{z} - \frac{eA}{v_{\Delta}}\right] \tau_{y} \right\}.
\end{align}

\enni Note the opposite sign for the vector potential wrt valley $(1)$, in accordance with time-reversal symmetry. From Ref.~\onlinecite{Balatsky}, we write the solutions as

\enni \begin{align}
\tilde{\Psi}^{(1')}_{s, n} = \text{sgn}(s)
\begin{pmatrix}
-i \alpha_{n} \psi_{n}\left[x, \lambda q_{z} \right] \\
\beta_{n} \psi_{n-1}\left[x, \lambda q_{z} \right]
\end{pmatrix},
\end{align}

\enni with 

\enni \begin{align}
\alpha_{n} = & \sqrt{\frac{E_{n} + v_{\Delta} \text{sgn}(s) q_{y} }{2E_{n}}} \\
\beta_{n} = & \sqrt{\frac{E_{n} - v_{\Delta} \text{sgn}(s) q_{y} }{2E_{n}}}.
\end{align}

In the un-rotated basis the solutions are

\enni \begin{align}
\Psi^{(1')}_{s, n} = 
\frac{\text{sgn}(s)}{\sqrt{2}}
\begin{pmatrix}
-i \alpha_{n} \psi_{n}\left[x, \lambda q_{z} \right] + i \beta_{n} \psi_{n-1}\left[x, \lambda q_{z} \right] \\
 \alpha_{n} \psi_{n}\left[x, \lambda q_{z} \right]
+ \beta_{n} \psi_{n}\left[x, \lambda q_{z} \right]
\end{pmatrix}.
\end{align}

\enni Formally, the solutions can be converted to the initial sign of the vector potential by sending $q_{z} \rightarrow -q_{z}$ and taking into account the parity of the Hermite polynomials.  

For the zeroth LL, $\beta_{0}=0$ and $\alpha_{0}=1$ and 
the coefficients are given by 

\enni \begin{align}
\Psi^{(1')}_{s, n} = 
\frac{\text{sgn}(s)\psi_{0}\left[x, -\lambda q_{z} \right]}{\sqrt{2}}
\begin{pmatrix}
-i   \\
 1
\end{pmatrix}, 
\end{align}

\enni with energies

\enni \begin{align}
E_{0}(q_{y}, q_{z}) = v_{\Delta} \text{sgn}(q_{y}) q_{y}.
\end{align}

\enni The zeroth LLs at valley $(1')$ have the same chiralities 
as their valley $(1)$ correspondents.  

The explicit zeroth LL solutions are 

\enni \begin{align}
\gamma^{(1)}_{\up, 0}(q_{y}, q_{z}) = & \int dx \frac{\psi_{0}(x, q_{z} )}{\sqrt{2}} \left[ i \Psi^{(1)}_{\up}(x, q_{y}, q_{z}) - \Psi^{(1'), \dag}_{\up}(x, -q_{y}, -q_{z}) \right], \label{Eq:Trpl_LL_1} \\
\gamma^{(1)}_{\dn, 0}(q_{y}, q_{z}) = & \int dx \frac{\psi_{0}(x, q_{z} )}{\sqrt{2}} \left[ -i \Psi^{(1)}_{\dn}(x, q_{y}, q_{z}) - \Psi^{(1'), \dag}_{\dn}(x, -q_{y}, -q_{z}) \right] \\
\gamma^{(1')}_{\up, 0}(q_{y}, q_{z}) = & \int dx \frac{\psi_{0}(x, -q_{z} )}{\sqrt{2}} \left[ i \Psi^{(1')}_{\up}(x, q_{y}, q_{z}) - \Psi^{(1), \dag}_{\up}(x, -q_{y}, -q{z}) \right] \\
\gamma^{(1')}_{\dn, 0}(q_{y}, q_{z}) = & \int dx \frac{\psi_{0}(x, -q_{z} )}{\sqrt{2}} \left[ -i \Psi^{(1')}_{\dn}(x, q_{y}, q_{z}) - \Psi^{(1), \dag}_{\dn}(x, -q_{y}, -q_{z}) \right]. \label{Eq:Trpl_LL_2}
\end{align}

As for the case of singlet pairing, only two out of the 
above four states are independent since 

\enni \begin{align}
\gamma^{(1')}_{\up, 0}(q_{y}, q_{z}) = & -i \gamma^{(1), \dag}_{\up, 0}(-q_{y}, -q_{z}) \\
\gamma^{(1')}_{\dn, 0}(q_{y}, q_{z}) = & i \gamma^{(1), \dag}_{\dn, 0}(-q_{y}, -q_{z}). 
\end{align}

\subsection{Projection of fields onto the zeroth LLs}
\label{Sec:IVc}

We follow the arguments of Sec.~\ref{Sec:IId} to obtain the Fermi fields via the inverse  BdG transformation:

\enni \begin{align}
\Psi^{(\alpha)}_{\sigma}\left(x, q \right)= \sum_{n, s} u^{(\alpha)}_{s\sigma, n}(x, q) \gamma^{(\alpha)}_{s, n}(k) + 
v^{(\bar{\alpha}),*}_{s \sigma, n}(x, -q) \gamma^{(\bar{\alpha}),\dagger}_{s, n}(-q).
\end{align} 

\enni In the spin-triplet case $u^{(\alpha)}_{s\sigma} = \delta_{s\sigma} u_{s}$ while $v^{(\alpha)}_{s\sigma}= (-\sigma_{z})_{s\sigma} v_{\bar{s}}$. Only two out of the four degrees-of-freedom in Eqs.~\ref{Eq:Trpl_LL_1}-\ref{Eq:Trpl_LL_2} are independent. Therefore we must truncate the Hilbert space of the zeroth LL to ensure a well-defined BdG transformation: 

\enni \begin{align}
\Psi^{(1)}_{P,\up}(x,\bm{q}) = & \frac{ -i\psi_{0}(x, q_{z})}{\sqrt{2}} \gamma^{(1)}_{\up, 0}(\bm{q}) \\
\Psi^{(1)}_{P,\dn}(x,\bm{q}) = & \frac{ i\psi_{0}(x, q_{z})}{\sqrt{2}} \gamma^{(1)}_{\dn, 0}(\bm{q}) \\
\Psi^{(1')}_{P,\up}(x, \bm{q}) = & \frac{-\psi_{0}(x, -q_{z})}{\sqrt{2}} \gamma^{(1), \dag}_{\dn, 0}(-\bm{q}) \\
\Psi^{(1')}_{P,\dn}(x, \bm{q}) = & \frac{-\psi_{0}(x, -q_{z})}{\sqrt{2}} \gamma^{(1), \dag}_{\up, 0}(-\bm{q}).
\end{align}

\enni In turn, the projections onto the zeroth LL are not independent:

\enni \begin{align}
\Psi^{(1')}_{P,\sigma}(x,q) = & i \text{sgn}(\sigma) \Psi^{(1), \dagger}_{P,\sigma}(x,-q). 
\end{align}

\enni As in the case of singlet pairing, these relations are matched by the real-space operators

\enni \begin{align}
\Psi^{(1')}_{P,\sigma}(x,y,z) = & \int \int \left( \frac{dq_{y}}{2\pi} \right) 
\left( \frac{dq_{z}}{2\pi} \right) e^{i q_{y}y} e^{i q_{z}z} \Psi^{(1')}_{P,\sigma}(x,q_{y},q_{z}) \notag \\
= & \int \int \left( \frac{dq_{y}}{2\pi} \right) 
\left( \frac{dq_{z}}{2\pi} \right) e^{-i q_{y}y} e^{-i q_{z}z} \Psi^{(1')}_{P,\sigma}(x,-q_{y},-q_{z}) \notag \\
= & i \text{sgn}(\sigma) \int \int \left( \frac{dq_{y}}{2\pi} \right) 
\left( \frac{dq_{z}}{2\pi} \right) e^{-i q_{y}y} e^{-i q_{z}z} \Psi^{(1), \dag}_{P,\sigma}(x,q_{y},q_{z}) \notag \\
= & i \text{sgn}(\sigma)\Psi^{(1), \dag}_{P,\sigma}(x,y,z).
\end{align} 

\enni We can map the valley $(1')$ operators to those at valley $(1)$. From Sec.~\ref{Sec:IId}, these satisfy 

\enni \begin{align}
\{ \Psi^{(1)}_{\sigma}(\bm{r}), \Psi^{(1), \dagger}_{\sigma'}(\bm{r}') \} = 
\approx & \delta_{\sigma, \sigma'}\frac{1}{2 \lambda l^{2}_{B}} e^{-\frac{( x - x')^{2}}{4 \lambda^{2} l^{2}_{B}}} e^{-\frac{(z - z')^{2}}{4 \lambda^{2} l^{2}_{B}}} e^{\frac{i (x+x')(z-z')}{2 \lambda l^{2}_{B}}} \delta(y-y')
\end{align}

\enni \begin{align}
\{ \Psi^{(1)}_{\sigma}, \Psi^{(1)}_{\sigma'} \} = & 0 \\
\{ \Psi^{(1), \dag}_{\sigma}, \Psi^{(1), \dag}_{\sigma'} \} = & 0.
\end{align}

The projection of the field $\Psi_{\sigma}(\bm{r})$ with contributions from both valleys, onto the zeroth LL is given by 

\enni \begin{align}
\Psi_{P,\sigma}(\bm{r}) = & e^{iK_{Fz}z} \Psi^{(1)}_{P,\sigma}(\bm{r}) 
+ e^{-iK_{Fz}z} \Psi^{(1')}_{P,\sigma}(\bm{r}) \notag \\
= & e^{iK_{Fz}z} \Psi^{(1)}_{P,\sigma}(\bm{r}) 
+ i \text{sgn}(\sigma) e^{-iK_{Fz}z} \Psi^{(1), \dag}_{P,\sigma}(\bm{r}).
\end{align}

\enni We deduce that 

\enni \begin{align}
\Psi^{\dag}_{P,\sigma}(\bm{r}) = & e^{-iK_{Fz}z} \Psi^{(1), \dag}_{P,\sigma}(\bm{r}) 
- i \text{sgn}(\sigma) e^{iK_{Fz}z} \Psi^{(1)}_{P,\sigma}(\bm{r}) \notag \\
= & - i \text{sgn}(\sigma) \Psi_{P,\sigma}(\bm{r}).
\label{Eq:Trpl_Mrn_ancm}
\end{align}

\enni This relation clearly signals emergent real-space Majorana fermions. Indeed, these projected fields obey 

\enni \begin{align}
\left\{ \Psi_{P,\sigma}(\bm{r}), \Psi_{P,\sigma'}(\bm{r}') \right\} 
= & \cos\left(\left[K_{F} + \frac{(x+x')}{2 \lambda l^{2}_{B}} \right] (z-z') \right) \left[ \frac{1}{ \sqrt{\lambda} l_{B}}  e^{-\frac{( x - x')^{2}}{2 \lambda^{2} l^{2}_{B}}} \right] \left[ \frac{1}{ \sqrt{\lambda} l_{B}} e^{-\frac{(z - z')^{2}}{4 \lambda^{2} l^{2}_{B}}} \right] \delta_{\sigma, \sigma'} \delta(y-y').
\label{Eq:Trpl_rl_spc_Mjrn}
\end{align}

\enni The first term represents fast oscillations. We recognize the remaining terms as the contributions from the zeroth LL harmonic oscillator wavefunctions. The last term is a Dirac delta function which is typical of free Majorana fields~\cite{Elliot}. 

The real-space Majorana nature of the projected fields leading to Eq.~\ref{Eq:Trpl_rl_spc_Mjrn} was illustrated in the continuum. However, this property is not intrinsic to this limit. Indeed, it follows from the general p-h symmetry and topology of the bulk Hamiltonian under strain as illustrated by the general arguments of Sec.~\ref{Sec:I}. We can therefore extend the Majorana nature to operators defined on the lattice such that $c^{\dag}_{P, \sigma}(\bm{R})= -i \text{sgn}(\sigma)c_{P, \bar{\sigma}}(\bm{R})$, as mentioned in the main text. These are expected to obey a lattice version of Eq.~\ref{Eq:Trpl_Mrn_ancm}.

\section{Estimates of strain}
\label{Sec:V}

\subsection{General Formulae}
\label{Sec:Va}

In the continuum limit, the effect of uniaxial strain of the hopping coefficients of our models is given by~\cite{Nica, Castro}

\enni \begin{align}
\delta t(x) = t \beta u_{xx},
\end{align}

\enni where $t$ is the pristine value of the hopping. $\beta$ is given by

\enni \begin{align}
\beta = \frac{d \ln t}{\ ln a},
\end{align}

\enni where $a$ is the nearest-neighbor distance. 
It is typically associated with the Gr\"{u}neisen parameter~\cite{Vozmediano}. $u_{xx}$ is a diagonal component of the strain tensor.

In the low-energy theory, $\delta t(x)$ determines an effective vector potential via 

\enni \begin{align}
\tilde{A}_{y}(x) = & \frac{\delta t(x)}{e} \notag \\
= & \frac{t \beta u_{xx}(x)}{e},
\end{align}

\enni where $e$ is the bare electronic charge. 
For illustration purposes, we chose the vector potential along the $y$ direction, but generalization to other cases of uniaxial strain are obvious. 

In order to produce an approximately uniform pseudo-magnetic field, we require

\enni \begin{align}
\partial_{x} \tilde{A}_{y}(x) = & \left( \frac{t\beta}{e} \right) \partial_{x} u_{xx} \notag \\
= & B_{z}.
\end{align}

From the previous expression, it is clear that we require 
a strain which, to leading order, varies linearly with distance in the bulk of the sample:

\enni \begin{align}
u_{xx} \approx \rho x,
\end{align}

\enni where $\rho$ is the gradient of the strain. It is convenient to define a dimensionless parameter 

\enni \begin{align}
\tilde{\rho}= \rho a.
\end{align}

\enni Note that the strain parameter $\epsilon$ discussed 
in the main text is given by 

\enni \begin{align}
\epsilon= \beta \rho.
\end{align}

All of the relevant scales of the problems can be written 
in terms of the parameters $t, \Delta, \beta, \tilde{\rho}$, representing hopping coefficient, pairing amplitude, Gr\"{u}neisen parameter, and rate of strain increase per unit cell, respectively. The pseudo-magnetic length is determined by 

\enni \begin{align}
l_{B} = & \sqrt{\frac{v_{\Delta}}{e B}} \notag \\
\approx & a \sqrt{\left(\frac{\Delta}{t} \right) \left(\frac{1}{\beta \tilde{\rho}} \right)}.
\end{align}

\enni The LL separation in energy is given by 

\enni \begin{align}
E_{c} = & \hbar \omega_{c} \notag \\
= & \sqrt{2} \frac{\Delta a}{l_{B}} \notag \\
= & \Delta \sqrt{\frac{2 \beta \tilde{\rho}}{\Delta/t}},
\end{align}

\enni where we re-introduced factors of $\hbar$ in Eq.~\ref{Eq:Sngl_Frmi_vlct} with $v_{\Delta} \approx \Delta a/ \hbar$.

\subsection{Numerical Estimates}
\label{Sec:Vb}

Using the formulae of the previous section, we can estimate typical pseudo-magnetic lengths and LL separation in energy using parameters for high-$T_{c}$ materials: 
$\Delta \approx$30 meV~\cite{Hashimoto}, $a \approx$  3.9 $\AA$, $t \approx$ 0.38 eV~\cite{Korshunov}, and $\beta \approx 1.2$~\cite{Mouallem}.

Considering a maximum strain of 0.05 achieved over a typical length scale approximately $L=$0.05 mm, we can estimate $\tilde{\rho}$ as

\enni \begin{align}
\tilde{\rho} = & 0.05 \left(\frac{a}{L} \right) \notag \\
= & 3.9 \times 10^{-7}.
\end{align}

\enni This gives 

\enni \begin{align}
l_{B} \approx &~410 a \\
E_{c} \approx &~0.11~\text{meV}~(1.2~\text{K}).
\end{align}

\end{document}